\documentclass[]{interact}

\usepackage{epstopdf}
\usepackage[caption=false]{subfig}

\usepackage[natbibapa,nodoi]{apacite}
\theoremstyle{plain}

\theoremstyle{definition}

\theoremstyle{remark}

\usepackage{amsmath,amssymb,amsthm}
\usepackage{thmtools}
\newtheorem{hyp}{Sub-hypothesis}
\usepackage{hyperref}
\usepackage{footmisc}
\usepackage[table]{xcolor}
\usepackage{graphicx}
\usepackage{booktabs}
\usepackage{algorithm}
\usepackage{algpseudocode}
\usepackage{listings}
\usepackage{tabularx}

\begin{document}

\articletype{ARTICLE TEMPLATE}

\title{Templates of generic geographic information for answering where-questions}

\author{
\name{Ehsan Hamzei\thanks{Correspondence concerning this article should be addressed to Ehsan Hamzei, Department of Infrastructure Engineering, The University of Melbourne, Parkville, VIC 3010, Australia; Email: ehamzei@student.unimelb.edu.au}, Stephan Winter and Martin Tomko\textsuperscript{a}}
\affil{\textsuperscript{a}Department of Infrastructure Engineering, The University of Melbourne, Parkville, Australia}
}

\maketitle

\begin{abstract}
In everyday communication, where-questions are answered by place descriptions. To answer where-questions automatically, computers should be able to generate relevant place descriptions that satisfy inquirers' information needs. Human-generated answers to where-questions constructed based on a few anchor places that characterize the location of inquired places. The challenge for automatically generating such relevant responses stems from \textit{selecting relevant anchor places}. In this paper, we present \textit{templates} that allow to characterize the human-generated answers and to imitate their structure. These templates are patterns of generic geographic information derived and encoded from the largest available machine comprehension dataset, MS MARCO v2.1. In our approach, the toponyms in the questions and answers of the dataset are encoded into sequences of generic information. Next, sequence prediction methods are used to model the relation between the generic information in the questions and their answers. Finally, we evaluate the performance of predicting templates for answers to where-questions.
\end{abstract}


\section{Introduction}
Consider the following question and its corresponding answer, taken from the Microsoft Machine Comprehension (MS MARCO) dataset v2.1 \citep{Nguyen2016ms}:

\vspace{0.2cm}
\textbf{Question}: \textit{Where is Putney Bridge?}

\textbf{Answer}: \textit{Putney Bridge is a bridge crossing of the River Thames in west London.}
\vspace{0.2cm}

\noindent This \textit{where}-question is answered using a \emph{place description} -- a description that characterizes the location of interest (Putney Bridge) based on a few anchor places (River Thames and London). Place descriptions, however, are not the only way to answer where-questions. Where-questions can also be answered via other representations such as maps or sketches \citep{mapvstext:2010}. Invariant to the chosen representation, the answers localize the place in question based on its spatial relationships with chosen anchor places \citep{COUCLELIS198799}. Hence, answering where-questions poses the following challenges no matter what representation is used:
\begin{itemize}
    \item Generating informative answers -- i.e., the answer should complete the inquirers' gap of knowledge in a way that obvious or already-known responses should be avoided and useful and necessary information are included \citep{shanon1983answers}. In the example, obvious, inadequate or undetailed answers such as \textit{on Earth} or \textit{in the UK} or \textit{over a river} are avoided by the responder.
    
    \item Answering the question in a cognitively efficient manner \citep{wilson2002relevance} -- e.g., producing short and straightforward place descriptions \citep{agilePaper} and personalized map labelling strategies in map visualizations \citep{wilson2018systems}. In the example, the responder excludes unnecessary information such as the nearby theaters and restaurants to keep the answer as simple and relevant as possible.

    \item Determining the level of granularity of answers -- e.g., a suitable zoom level for maps \citep{ballatore2019} and referring to places of suitable granularity in place descriptions \citep{hamzeiCOSIT}. In our example, the name of the roads and streets that are connected to the bridge are neglected in the answer based on the judgement of the responder for the relevant scale level.
    
    \item Selecting places that can be assumed to be known by the inquirer -- e.g., labelling the places known to inquirers in maps \citep{suomela2009displaying} and referring to them in place descriptions as anchors. In the example, the location of \emph{River Thames} and \emph{London} are assumed to be known to the inquirer.
\end{itemize} 
Where these challenges are met by an answer, the communication succeeds. 

Addressing these challenges is a necessary step towards answering where-questions. To understand and imitate human selectivity in choosing anchor places, we investigate and characterize human-generated answers to where-questions. The results of our research are applied for generating answers to where-questions as natural language responses (place descriptions). Selecting relevant anchor places is an essential part of generating place descriptions that succeed in answering where-questions. Moreover, information about anchor places can be used in static maps to be visualized in a proper context frame \citep{ballatore2019}. 

Current geographic question answering systems are often focused on coordinate retrieval as answers to where-questions \citep[e.g.,][]{luque:2006, Stadler:2012}. While coordinates are useful for communication between location-based services to perform spatial analysis or visualization \citep{JIANG2006712}, it is not necessarily a relevant response to inquirers without a proper map visualization. Yet, a characterization of relevant anchor places to localize a place in question is still missing.

In this paper, we study human-generated answers to where-questions to inform the properties of such answers and to devise and test a method to imitate their structures in machine-generated responses. To achieve these goals, the information in a where-question and its answer is modelled as \textit{an ordered set of places} that are mentioned in their content. Then the properties of places in questions and corresponding answers are derived and further investigated. This model forms a template (i.e., an ordered set of place properties) that enables computers to learn and imitate human answering behaviour. In other words, place properties are utilized to understand why a set of places are chosen as anchors to localize the place in question and how this selectivity can be imitated by computers.

The properties that are used in the templates are generic geographic information that describe the shared meaning of places in form of generic types from a finite set of categories. Referring to the example above, the place in question is a \textit{bridge} which is localized by referring to the river it goes over and the city it belongs to. Here, the template captures the structure of the answer as relationships between \textit{bridges and rivers}, and \textit{bridges and cities}.

\subsection{Background: Geographic Question Answering}
Geographic Question Answering (GeoQA) is defined as methods and algorithms that help inquirers to satisfy their information need by deriving answers to their geographic questions. In GeoQA, answering geographic questions can be based on diverse information sources such as textual information \citep{Mishra:2010, Ferres:2006}, geodatabases \citep{chen2014}, and spatially-enabled knowledge bases \citep{Ferres:2010}. GeoQA (and in general QA) architectures typically resolve three tasks: (a) question classification and intent analysis, (b) finding relevant sources, and (c) extracting answers from the sources \citep{Ferres:2006}.

The classification of the questions \citep{agilePaper, Mohasseb2018} enables GeoQA to coarsely identify the intent and purpose of asking questions (e.g., localization, or navigation). Next, the questions are translated into formal representations such as database queries or even just a vector representation of extracted keywords \citep{Punjani:2018:TQA:3281354.3281362}. Using the representations, the information sources can be searched or queried to look up the possible answers \citep{Zheng2019}. Finally, the factoid answers are retrieved from the sources -- e.g., a sentence in a Web document, a cell in a selected table, or a node in a graph knowledge base \citep{sun-etal-2018-open}.

In recent years, several GeoQA studies were conducted for answering geographic questions \citep{Stadler:2012}, creating knowledge bases from unstructured data \citep{Mai:2018}, and relaxing unanswerable questions \citep{Mai:2020}. Focusing on answering geographic questions, previous studies provide solutions to retrieve responses from knowledge bases \citep{Stadler:2012} and documents \citep{Buscaldi:2006,luque:2006}. GeoQA studies are mostly focused on what/which questions about geographic places \citep[e.g.,][]{Scheider:2020, Vahedi:2016}. In answering where-questions, the task is either simplified into retrieving stored coordinates \citep{luque:2006, Stadler:2012}, or selecting a part of text without explicit adaptation to the question \citep{Buscaldi:2006}.

When answering where-questions, the answer extraction step is particularly challenging. Without a well-designed approach to imitate human answering behavior, the extracted answers can easily be over-specified and consequently uninterpretable for the inquirer, or under-specified and thus obvious and uninformative to the inquirer \citep{shanon1983answers}. Hence, the challenge is to provide relevant answers by selecting proper set of anchor places to localize the place in question.

\subsection{Rationale and Research Gap}
To enable computers to provide responses with similar qualities to human-generated answers, the responses need to be relevant. An answer is relevant if its positive cognitive effects to inquirers are large and the processing effort to achieve the effect is small \citep{wilson2002relevance}. In other words, answers should be informative enough and as straightforward as possible. Assuming human-generated answers are relevant responses, machine-generated responses should imitate the selectivity in human-generated answer to provide useful pieces of information and avoid unnecessary ones. Generating such relevant responses is the major prerequisite of intelligent GeoQA as defined by \cite{winter2009spatial}.

Generic information captures shared meaning of geographic places. While generic geographic information is not used in QA, it has been used to investigate and characterize place descriptions \citep{Richter_et_al_2013,Purves2007}, route descriptions \citep{Raubal:2002}, and regions \citep{tomko2008categorical}. 

This research hypothesizes that, at least in the English language, generic geographic information can be used to characterize human answering behavior and ultimately to generate templates for answering where-questions. We approach this hypothesis by addressing three sub-hypotheses.

\setcounter{hyp}{0}
\begin{hyp}[Characteristics of the answers] \label{hyp:a}
Human-generated answers to where-questions have special characteristics that can be described and characterized in terms of generic geographic information such as type, scale, and prominence;
\end{hyp}
 
\begin{hyp}[Relation between where-questions and their answers] \label{hyp:b}
There is a strong relationship between generic information in the content of where-questions and their answers which can be used to characterize human answering behavior;
\end{hyp}

\begin{hyp} [Generating answers to where-questions] \label{hyp:c}
If Hypotheses 1 and 2 hold, the characteristics of human-generated answers and the relation between the questions and their answers can be used to generate templates to answer to where-questions.
\end{hyp}

\noindent To investigate the hypotheses, the following research questions will be addressed:
\begin{enumerate}
    \item How can the characterizing patterns of the human-generated answers be derived?
    \item How does generic geographic information in where-questions relate to the generic information in their human-generated answers?
    \item How can the templates be generated to imitate the structure of human-generated answers?
\end{enumerate}
By addressing the research questions, we contribute:
\begin{itemize}
    \item A generalized approach to investigate human answering behavior to where-questions using generic geographic information;
    \item An investigation of the human-generated answers to where-question asked in Web search, using patterns of type, scale and prominence of places.
\end{itemize}

\section{Methodology}

To investigate the hypotheses, we propose a generalized approach of \emph{specific-generic translation}. Next, a method using specific-generic translation is devised to investigate the QA in interaction of people with a general-purpose search engine. Other QA scenarios (e.g., human-human dialogue, human-social bot interaction) may require different design of specific-generic translations.

\subsection{Specific-Generic Translation}

\begin{figure}
    \centering
    \includegraphics[width=0.95\textwidth]{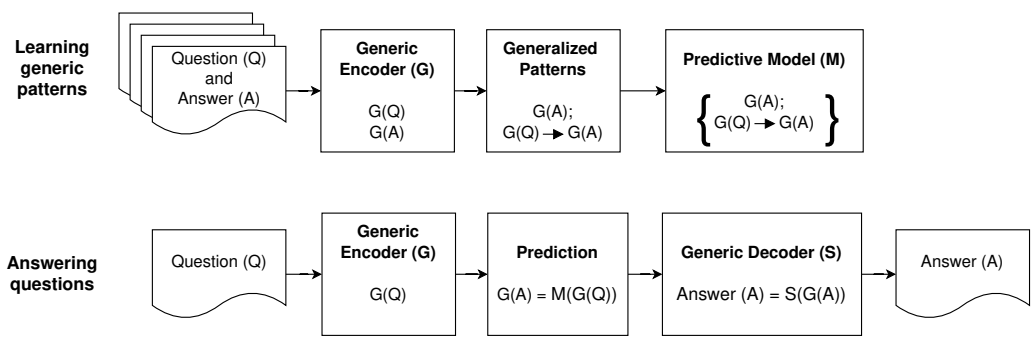}
    \caption{Specific-generic translation approach}
    \label{fig:method}
\end{figure}

Figure~\ref{fig:method} shows the proposed approach to derive characterizing patterns in human-generated answers and generating template to answer where-questions. The approach includes two stages: (1) \emph{learning generic patterns} where the objective is to investigate and to characterize human answering behavior into a machine learning model, and (2) \emph{answering questions} where the model is used to generate answers. The novelty of the approach is in the encoding of questions and answers into their generic meaning and to model the relation between questions and answers in their generic forms. Later, the model is used to generate generic forms of answers to where-questions. Finally, the answer is constructed by decoding generic form (e.g., type) of the answer into its specific representation (i.e., toponym).

The specific-generic translation approach involves the following steps:

\begin{enumerate}
    \item Selecting a set of generic information classes (e.g., place type, scale, prominence, functions and access) based on the context of QA and availability of data;
    
    \item Defining a schema for each selected generic information class;
    
    \item Designing an information extraction approach to encode the questions and answers into generic forms (\textit{Generic Encoder} in Figure~\ref{fig:method});
    
    \item Evaluating how effective each generic class is in capturing the relation between the questions and their human-generated answers (\textit{Generalized Patterns} in Figure~\ref{fig:method}). The results of evaluation also provide insights about human answering behavior in the context of the QA problem.
    
    \item Training a predictive model that can learn generalized patterns of human-generated answers (\textit{Predictive Model} in Figure~\ref{fig:method});
    
    \item Defining a decoding approach to map generic forms of answers into specific (toponym) representation (\textit{Generic Decoder} in Figure~\ref{fig:method}). This step can be followed by additional steps such as natural language generation to be used in real-world applications.
\end{enumerate}

In this paper, we discussed the results of the first five steps for question answering in a Web search scenario in details. The last step is only demonstrated using examples.

\subsection{Type, Scale and Prominence (TSP) Encoding}

Based on the specific-generic translation, TSP encoding is proposed to investigate where-questions constructed only with toponyms. The generic forms which are used to investigate these questions and their answers are \emph{type}, \emph{scale} and \emph{prominence} of the toponyms. We first introduce our terminology before discussing the details of the proposed TSP encoding method.

\subsubsection{Definitions}

The investigated types of where-questions are defined as:

\begin{itemize}
    \item \textbf{Toponym-Based Where-Question (TWQ)}: A toponym-based where-question is a geographical where-question that is generated completely using toponyms. For example, \emph{Where is Beverly Hills?} is a toponym-based where-question, while \emph{Where is Clueless filmed?} (without toponym) and \emph{Where can I buy furniture?} (affordance, buying furniture) do not belong to this type.

    \item \textbf{Simple Where-Question (SWQ)}: Simple where-questions are a sub-class of TWQs that contains only one toponym in their body (e.g., \emph{Where is Beverly Hills?}).

    \item \textbf{Detailed Where-Question (DWQ)}: Detailed where-questions are a sub-class of TWQs with more than one toponym in their content (e.g., \emph{Where is Beverly Hills, California?}). In DWQs, contextual details are provided in the content of the questions that shows what the inquirer already know -- e.g., Beverly Hills is located in California.
\end{itemize}

We use \emph{type}, \emph{scale} and \emph{prominence}, defined as:

\begin{itemize}
    \item \textbf{Type}: A type (e.g., restaurant, mountain) is a reference to a group of places with similar characteristics (e.g., affordance, or physical properties). Type defines similar places and differentiates dissimilar ones, sometimes in a hierarchical or taxonomical manner. Here, the relation between a specific reference to a place (unambiguous toponym) and its type is considered as a one-to-one relation.

    \item \textbf{Scale}: Scale is defined as a finite hierarchically-organized ordinal set of levels grounded in the human understanding and differentiation of places based on their size and their relationships (i.e., containment). The relation between scale and an unambiguous toponym is considered as one-to-one. Due to the specific context of the QA scenario, very fine levels of scale of geographic entities, such as room-level or object-level, can be neglected here, while in everyday human-human communication these levels of scale may have a more important role.

    \item \textbf{Prominence}: Prominence is a measure of how well-known a place is. In this research, prominence is characterized by a finite ordinal set of levels. While prominence of places is subjective and differs from person to person based on their experience, here prominence is considered as an absolute and objective measure to rank places, established through a proxy measure defined later. This approach enables to avoid individual experiential biases and is supported by the evidence of success in day to day communication in which the absolute prominence evaluation is adapted between hearers and speakers.
\end{itemize}

Type, scale and prominence are used to characterize place descriptions \citep{Richter_et_al_2013,Purves2007}. These geographic concepts can be used to capture different kinds of relationships among places. These relationships can be used to understand the relation between where-questions and their answers. For example, such relationships between rivers and seas (\textit{flows to}), and cities and countries (\textit{part of}) can be captured using place type. Considering the relation between where-questions and their answers, \textit{containment} (different levels) and \textit{nearness} (a same level) can be captured through differences among scale levels. Finally, prominence is a measure to check whether the answer is interpretable by the inquirers -- i.e., more prominent places are expected to be better known by inquires.

Finally, aspects of human-generated answers which are investigated in this paper are defined below:

\begin{itemize}
    \item \textbf{Content}: The content of an answer is a collection of distinct information units that are presented to satisfy the inquirer's information need. Content can be generic (e.g., type) or specific (e.g., toponym). Content is the most important aspect of the answers, in a way that the difference between correct and incorrect responses are completely based on their content.
    
    \item \textbf{Style}: The style of an answer is the way that the content is presented. Style directly influences the perception of naturalness of the response. Referring to the introductory example, \textit{\ldots Putney Bridge is a bridge crossing of the River Thames in west London} and \textit{\ldots Putney Bridge is a bridge in west London which goes over River Thames} are two different styles of answers (with same content) to the question. Here, the former is preferred by the responder.
\end{itemize}

\subsubsection{TSP Sequences}
In TSP encoding, we use a sequence representation to model generic/specific information in the questions and answers. A sequence is defined as an ordered set of items (here, references to generic/specific geographic information). We first model questions and their corresponding answers as sequences of toponyms (specific representation). Then, these toponym sequences are encoded into type, scale and prominence sequences by translating each specific toponym into its corresponding generic type, scale and prominence reference. Referring to the introductory example, the specific representations (toponym sequences) and the encoded type sequences (an example of a generic sequence) of question and answer are presented below:
\begin{itemize}
    \item \textbf{Toponym sequences}: [Putney Bridge] [River Thames, London]
    
    \item \textbf{Type sequences}: [bridge] [river, city]
\end{itemize}
Here, the \emph{content} refers to the information items in the sequences, and their order defines the \emph{style} in which the information is presented.

\section{Implementation}

\begin{figure}
    \centering
    \includegraphics[width=0.95\textwidth]{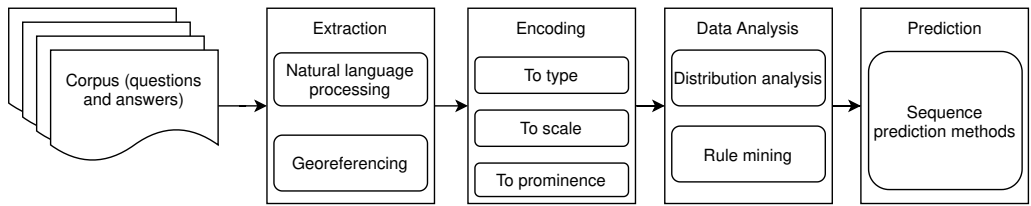}
    \caption{The proposed implementation approach}
    \label{fig:method2}
\end{figure}

Figure~\ref{fig:method2} shows the proposed workflow\footnote{Additional details of implementation are presented in the supplementary material (Section 1)} to investigate TWQs and their answers. Here, we detail the dataset, extraction, encoding, generic patterns and prediction. A complete implementation of the proposed TSP encoding approach also includes decoding from generic to specific information. Here, the decoding step is demonstrated through examples, and a fully automated implementation remains out of scope of this paper.

\subsection{Data}

The questions in MS MARCO v2.1 \citep{Nguyen2016ms} are categorized into five categories using tags: (1) \emph{numeric}, (2) \emph{entity}, (3) \emph{location}, (4) \emph{person}, and (5) \emph{description} \citep{Nguyen2016ms}. Geographic questions can thus be easily extracted using the predefined \emph{location} tag. The dataset contains over one million records divided into \emph{training}, \emph{development} and \emph{testing} subsets. 

Each record in the dataset includes a \emph{question}, \emph{human-generated answer(s)} (except for records in the \emph{test} dataset, where the answers are deliberately excluded), a \emph{tag}, and a \emph{set of documents} retrieved by the Microsoft Bing search engine\footnote{More information about the dataset can be found in: \url{https://github.com/dfcf93/MSMARCO}}. 

The `location questions' in MS MARCO (56,721 question-answer pairs) include 36,939 geographic questions, and the remainder are questions about fictional, mystic and other non-geographic places \citep{agilePaper}. Among the geographic questions, 13,195 pairs of questions and answers are geographic where-questions \citep{agilePaper}. 

There are several reasons to choose MS MARCO for this study considering other available datasets such as SQuAD \citep{rajpurkar2016squad}:
\begin{itemize}
    \item MS MARCO is the largest available QA dataset;
    
    \item The questions are labelled and geographic questions can be easily extracted;
    
    \item All questions are asked in a specific real-world scenario (i.e., Web search);
    
    \item Inquirers pose questions to resolve their information needs while in some datasets such as SQuAD, questions are made from documents. In other words, questions in SQuAD is more about what a document can answer rather than what actual inquirers want to know.
    
    \item The answers are provided using an open form strategy. The answerers can utilize suggested documents (one or more) and their own knowledge to answer a question. Hence, the answers are not directly extracted as a single span of a document.
\end{itemize}

\subsection{Extraction}

We first extract the questions labelled as \emph{location} and starting with a \emph{where}-token. Next, the text is the toponyms inside the questions and answers are identified using Named Entity Recognition (NER) and gazetteer lookups using both OSM Nominatim and Geonames gazetteers. Here, the Stanford NLP toolkit is used to extract named entities \citep{finkel2005incorporating}. In this step, if a compound noun phrase is tagged as location, first the compound noun is checked by gazetteer lookup; if it is not identified, then its constituent simple nouns are checked. If a compound or simple noun phrase is found in both gazetteers, it is stored as a toponym. For the extracted toponyms, we retain only records for which (1) the OSM and Geonames records have same name, and (2) the Geonames' point-based coordinates are inside the region-based representation of their corresponding OSM records.

The toponym disambiguation is then undertaken based on the \emph{minimum spatial context} heuristic \citep{Leidner:2003:GSN:1119394.1119399}. We use bounding boxes to determine which combination of the geographic locations satisfy the minimum spatial extent condition. In cases of duplicate places in GeoNames which lead to the same bounding boxes, the combination with more specific place types is selected. For example, populate place (PPL) is a place type in GeoNames which could be a village, city, state and even country. Hence, administrative divisions (e.g., state) are chosen over the populated places. Finally, if the toponym ambiguity still exists, we use importance value to select the more prominent combination.

More sophisticated heuristics in toponym disambiguation \citep[e.g.,][]{wang:2010, Lieberman:2012} are not used due to reliance on significant assumptions -- e.g., the relation between place types in the toponyms, city-country relation. These heuristics constrain the relationships between type, scale and prominence of resolved places in the text. This may impact the results of this study and lead to stronger associations based on type, scale and prominence between toponyms in questions and answers. Here, to present fair results, we avoid using these disambiguation methods.

\subsection{Encoding}

The gazetteers' attributes for the extracted toponyms have been used as proxies to capture type, scale and prominence of the toponyms in questions and answers. Using these proxies, the sequence representations for each question-answer pair are encoded into TSP sequences.

 \textbf{Type:} The Geonames type schema\footnote{\url{https://www.geonames.org/export/codes.html}\label{note1}} has been used without modification to encode generic place types. This schema contains 667 unique types of geographic features, covering both natural and man-made geographic types.

 \textbf{Scale:} To capture scale, we have extended the schema from \cite{Richter_et_al_2013}. This schema contains seven levels of granularity: (1) furniture, (2) room, (3) building, (4) street, (5) district, (6) city, and (7) country. We have extended the coarse levels of granularity by adding \emph{county}, \emph{state}, \emph{country}, and \emph{continent}, and removing the \emph{furniture} and \emph{room} levels from the schema. OSM records include an attribute related to the OSM definition of scale (i.e., place\_rank\footnote{\url{https://wiki.openstreetmap.org/wiki/Nominatim/Development\_overview}\label{note2}}), a number between 0 to 30. We convert the extracted gazetteers' records into the appropriate scale level based on a look-up table that maps OSM scale levels into the proposed scale schema (see the supplementary material, Section 1.1).

 \textbf{Prominence:} To capture prominence, the \emph{importance} attribute in the extracted OSM Nominatim record is used. The OSM importance value is estimated using Wikipedia importance score \citep{thalhammer2016pagerank} with some minor tweaks\footnote{\url{https://lists.openstreetmap.org/pipermail/geocoding/2013-August/000916.html}}. The value is defined between 0 and 1, and it is designed to be used for ranking search results. We translate these values into seven finite levels of prominence, derived by \emph{natural breaks} classification \citep{Jensk1967IYC} of the frequency spectrum of the values.

\subsection{Distribution Analysis and Rule Mining}

Distribution analysis and rule mining techniques are used to extract and investigate patterns in the human-generated answers and the relation between the questions and their answers. Distributions of type, scale and prominence sequences are used to compare the questions and answers. To derive patterns in the questions and their answers, association rule mining, a-priori algorithm \citep{Agrawal:1994}, is used. 

The strength of the extracted rules are evaluated using the standard measures -- i.e., \emph{support}, \emph{confidence}, and \emph{lift}. Support defines how frequently an association rule is observed in the whole dataset, and confidence determines how often the rule is true. Lift is a measure to evaluate the importance of the rules -- i.e., lift greater than one shows positive and strong dependency among the elements of the extracted rule. This part of the method is devised to test the first and second hypotheses.

\subsection{Prediction}

The input for the prediction is an encoded sequence of TWQs, and the output is the generic sequence of their corresponding answers. The problem can then be formulated as a sequence prediction from concatenated generic sequences for the questions and their answers, where a part of a sequence is known, and the rest is predicted. Table~\ref{tab:seqMethods} shows the sequence prediction methods which are used in this study. We used and extended an open-source toolkit for sequence analysis \citep{SPMF} to implement the prediction methods.

These classic methods are divided into probabilistic \citep{cleary1984data,pitkow1999mininglongestrepeatin,padmanabhan1996using} and non-probabilistic categories \citep{ziv1978compression,laird1994discrete,gueniche2013compact,gueniche2015cptplus}. The probabilistic methods are based a graph representation of conditional probabilities \citep{cleary1984data} or Markov chain's transition probability matrix \citep{pitkow1999mininglongestrepeatin, padmanabhan1996using} of the sequence elements. The non-probabilistic methods compress the sequences in a lossy \citep{ziv1978compression, laird1994discrete} or lossless approaches \citep{gueniche2013compact, gueniche2015cptplus} into tree-based \citep{gueniche2013compact, gueniche2015cptplus} or graph-based \citep{laird1994discrete} data structures (for a review of sequence prediction methods see \cite{review:2020}).

The structure of sequence and the relation between prior elements in the sequence to their succeeding elements are trained into a model. The model is then tested on an unseen part of data using K-fold cross validation (K=10). We considered two baseline methods to evaluate the performance of the sequence prediction methods: (1) random sequence generation and (2) most frequent pattern. 

The random generation baseline only utilizes the schema of type, scale and prominence without any information about the distributions of values in the answers. The most frequent patterns baseline predicts templates of answers using the schema and the distribution of generic references in the answers. The difference between the prediction performances of random generation and the most frequent patterns shows the impacts of using the distribution of generic values in generating templates of answers (see hypothesis 1). The sequence prediction methods also consider the relation between generic values in the questions to their answers. Consequently, the improvement in generating the templates compared to the most frequent patterns baseline is related to the association between generic values of questions and their answers (hypothesis 2).

\begin{table}
\centering
	\caption{\label{tab:seqMethods}Sequence prediction methods}
	\resizebox{0.9\textwidth}{!}{
	\begin{tabular}{lll}
	\toprule
		\textbf{Method} & Publication & \textbf{Year}\\
	\midrule
        Lampel-Ziv 1978 (LZ78) & \citep{ziv1978compression}& 1978\\
        
        First order Markov Chains (Mark1) & \citep{cleary1984data}& 1984\\
        
        Transition Directed Acyclic Graph (TDAG) & \citep{laird1994discrete}& 1994\\
        
        Dependency Graph (DG) & \citep{padmanabhan1996using}& 1996\\
        
        All-k-Order Markov Chains (AKOM) & \citep{pitkow1999mininglongestrepeatin}&1999\\
        
        Compact Prediction Tree (CPT) & \citep{gueniche2013compact}&2013\\
        
        Compact Prediction Tree Plus (CPT+) &\citep{gueniche2015cptplus}&2015\\
    \bottomrule
	\end{tabular}
	}
\end{table}

In prediction, each generic form of questions is used to predict the same generic form of their answers. In addition, we have devised an approach to predict one of the generic forms of an answer using all generic forms (i.e., type, scale and prominence) of its corresponding question. Algorithm~\ref{alg:tsp} shows the process to use all three type/scale/prominence sequences to predict a generic form of the answers in each generic class. Here, each combination of type, scale and prominence values are mapped to a unique code. Using these codes, a new sequence is generated for each question/answer to capture type, scale and prominence together. Next, these sequences are used to predict the generic form of answers. Finally, a reverse mapping is used to decode these sequences into type, scale and prominence sequences.

\begin{algorithm}[ht]
\small
\caption{Training and prediction based on type-scale-prominence together}
\label{alg:tsp}
\begin{algorithmic}[1]
\Procedure{$\mathbf{TSP\_Prediction}$}{$type$, $scale$, $prominence$}
    \State generate a code for each unique combination of $type$-$scale$-$prominence$ ($TSP$)
    \State create encoded sequences based on generated $TSP$ codes
    \State train a model to predict $TSP$ in answers based on $TSP$ in the questions
    \For{every $question$}                    
        \State given a $question$ ($TSP$); predict the $answer$ ($TSP$)
        \State decode the predicted $answer$ ($TSP$) to $answer$ ($type$/$scale$/$prominence$)
        \If {multiple predictions are allowed}
            \State avoid counting duplicate decoded values for $type$/$scale$/$prominence$
        \EndIf
    \EndFor
\EndProcedure
\Statex
\end{algorithmic}
\end{algorithm}

\section{Results}

\subsection{Extraction and Encoding}

The assessment of toponym extraction, finding TWQs, and categorizing the questions into SWQs and DWQs are presented in Table~\ref{tab:extractionResults}. Here, average precision and recall of the extraction results are calculated using manually annotated data (5\% of TBWQs and their answers). For the task of finding TWQs in the dataset, the \emph{false negatives} (TWQs that have not been extracted) are not investigated, hence the recall is unknown. 

As shown in Table~\ref{tab:extractionResults}, 6,274 TWQs and their answers are found in the dataset. The TWQs are approximately 11.1\% of the \emph{location questions} of the dataset. For evaluation, 5\% of extracted TWQs (314 questions) are investigated and the precision of extraction is 91.7\% -- i.e., 288 of 314 extracted questions are TWQs. Using the 288 TWQs, the precision and recall of extracting toponyms and classifying the questions to SWQs and DWQs are presented in Table~\ref{tab:extractionResults}.

\begin{table}
\centering
	\caption{\label{tab:extractionResults}Extraction evaluation}
	\resizebox{\textwidth}{!}{
	\begin{tabular}{lllll}
	\toprule
		\textbf{Extraction} & \textbf{\#Extracted}& \textbf{\#Investigated} & \textbf{Precision} & \textbf{Recall} \\
	\midrule
       TWQs&6274&314 (5\%)&91.7\% (288 out of 314)&--\\
    \midrule
       SWQs&3285&121 out of 288&89.4\%&90.2\%\\
       DWQs&2989&167 out of 288&92.7\%&92.1\%\\
    \midrule
       Toponyms&22307&1133\footnote{unique toponyms extracted from the sampled questions and answers}&88.6\%&90.8\%\\
    \bottomrule
	\end{tabular}
	}
\end{table}

Table~\ref{tab:encodingResults} shows the number of records that are completely encoded for question-answer pairs in type, scale and prominence sequences. Here, if even the information for one place (which is mentioned either in the question or its answer) is missing, the question and its answer are not used to extract patterns or test the predictability of generating generic form the answer. As shown in the table, the encoding into scale and prominence is not always possible due to incompleteness of attribute information (i.e., \emph{place\_rank} and \emph{importance}) in OSM Nominatim.

\begin{table}
\centering
	\caption{\label{tab:encodingResults}Encoding results}
	\begin{tabular}{llll}
	\toprule
		\textbf{Encoding} & \textbf{\#TWQs}& \textbf{\#SWQs}&\textbf{\#DWQs}\\
	\midrule
       Type sequences        &6,274 & 3,285 & 2,989\\
       Scale sequences       &3,936 & 1,985 & 1,951\\
       Prominence sequences  &6,051 & 3,098 & 2,953\\
    \bottomrule
	\end{tabular}
\end{table}

\subsection{Distributions}

\begin{figure}
    \centering
    \includegraphics[width=\textwidth]{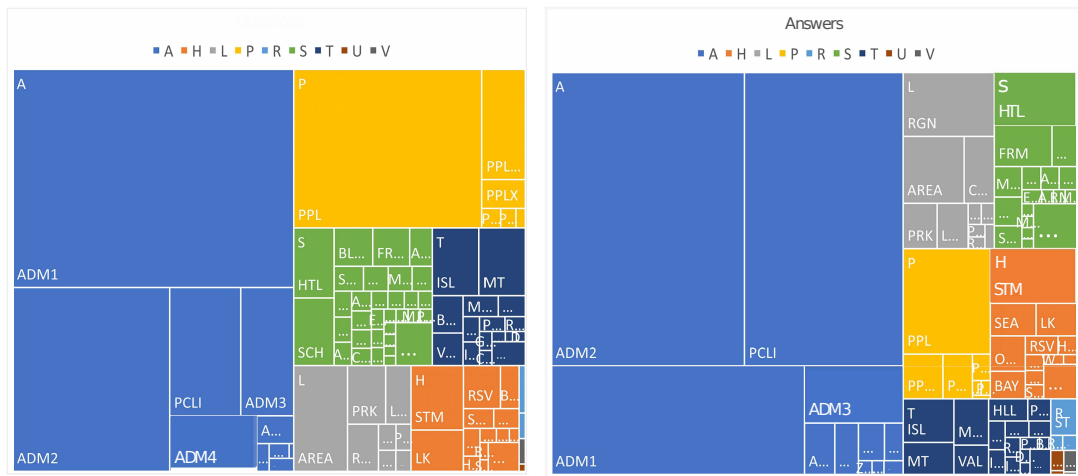}
    \caption{Distribution of place types in the questions and in the answers.}
    \label{fig:tqvsa}
\end{figure}

\begin{figure}
    \centering
    \includegraphics[width=\textwidth]{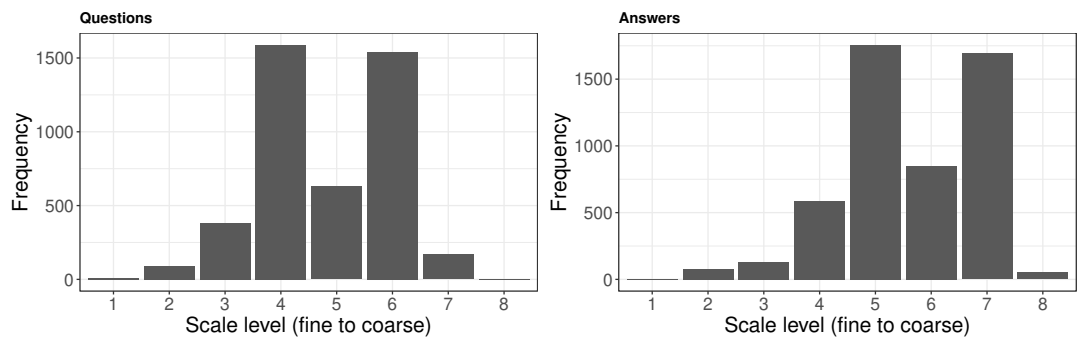}
    \caption{Distribution of levels of scale in all toponym-based where questions and answers.}
    \label{fig:sqvsa}
\end{figure}

\begin{figure}
    \centering
    \includegraphics[width=\textwidth]{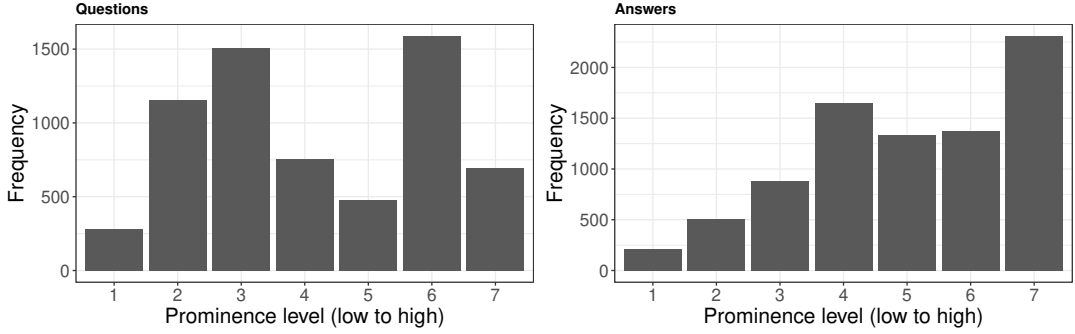}
    \caption{Distribution of prominence levels in the questions versus answers}
    \label{fig:pqvsa}
\end{figure}

The distribution of TWQs\footnote{A detailed comparison of SWQs and DWQs is presented in the supplementary material (Section 2)} and their answers based on type, scale and prominence are shown in Figures~\ref{fig:tqvsa},~\ref{fig:sqvsa} and~\ref{fig:pqvsa}. Figure~\ref{fig:tqvsa} shows that the diversity of types in the questions is higher than in the answers. While administrative divisions are more frequent than other generic types in both questions and answers, they are more dominant in the answers.

Figure~\ref{fig:sqvsa} shows the scale in the answers is systematically one level coarser than in the questions. In addition, the distribution shows that city-level and state-level scales are frequently observed in the questions, while the answers mostly contain references of county and country levels of scale. The results further show that the coarsest level of scale (i.e., continent level) is rarely observed in the answers. This observation shows an answer at the continent level would be under-specified in most cases, and therefore uninformative.

The distributions of prominence levels in questions and answers are similar to the distributions by scale (Figure~\ref{fig:pqvsa}). In the questions, we observe a bi-modal distribution of levels of prominence in the content of questions. The distribution of prominence in the answers, however, shows that higher levels are dominant. In contrast to the distributions by scale, the most prominent level is dominant in the answers. Hence, people tend to refer to well-known places in their answers. Unlike with scale, the highest levels of prominence do not necessarily lead to obvious or irrelevant answers\footnote{A detailed analysis of sequence distributions is available in supplementary material (Section 3)}.

\subsection{Extracted Rules}
To test Hypotheses 1 and 2 (see Section 1.4), we extract strong rules in the encoded pairs of questions-answers through association rule mining. The association rules extracted from the answers can be used to describe how answers are constructed in detail (Hypothesis 1). The relationship between the content of the questions and their answers can thus also be further investigated (Hypothesis 2).

Tables~\ref{tab:trules}-\ref{tab:prules} show the top five extracted rules (based on \emph{frequency}/\emph{support}) for type, scale and prominence, respectively. In the tables, the values starting with \emph{Q-} relate to the contents of the questions and the values starting with \emph{A-} to the content of the corresponding answers. As shown in the tables, some rules describe the structure of answers (e.g., \{A-ADM1, A-ADM2\} =\textgreater \{A-PCLI\}) while the others describe the relationships between questions and answers (e.g., \{Q-ADM2\} =\textgreater \{A-PCLI\}).

\begin{table}
    \centering
    \caption{\label{tab:trules}Extracted rules from type sequences}
    \resizebox{0.9\textwidth}{!}{
        \begin{tabular}{llllll}
        \toprule
            \textbf{Rank}&\textbf{rule}&\textbf{support}&\textbf{confidence}&\textbf{lift}&\textbf{frequency} \\
        \midrule
        \multicolumn{6}{c}{\textbf{Simple where-questions}}\\
        \midrule
            1 & \{A-ADM2\} =\textgreater \{A-ADM1\}        & 0.15    & 0.52       & 1.28 & 478   \\
            2 & \{Q-ADM1\} =\textgreater \{A-PCLI\}        & 0.08    & 0.74       & 1.69 & 259   \\
            3 & \{A-ADM1,A-ADM2\} =\textgreater \{A-PCLI\} & 0.08    & 0.54       & 1.24 & 259   \\
            4 & \{Q-ADM2\} =\textgreater \{A-PCLI\}        & 0.06    & 0.52       & 1.21 & 188   \\
            5 & \{Q-PPL,A-ADM2\} =\textgreater \{A-PCLI\}  & 0.04    & 0.54       & 1.23 & 112   \\
        \midrule
        \multicolumn{6}{c}{\textbf{Detailed where-questions}}\\
        \midrule
            1 & \{Q-ADM1\} =\textgreater \{A-ADM2\}        & 0.57    & 0.76       & 1.12 & 1701  \\
            2 & \{Q-ADM1\} =\textgreater \{A-PCLI\}        & 0.38    & 0.50       & 1.13 & 1126  \\
            3 & \{A-PCLI\} =\textgreater \{A-ADM2\}        & 0.35    & 0.79       & 1.17 & 1053  \\
            4 & \{A-ADM2,Q-ADM1\} =\textgreater \{A-PCLI\} & 0.31    & 0.54       & 1.21 & 916   \\
            5 & \{Q-PPL\} =\textgreater \{A-ADM2\}         & 0.22    & 0.78       & 1.15 & 656 \\
        \bottomrule
    \end{tabular}
    }
\end{table}

\begin{table}
    \centering
    \caption{\label{tab:srules}Extracted rules from scale sequences}
    \resizebox{0.75\textwidth}{!}{
        \begin{tabular}{llllll}
        \toprule
            \textbf{Rank}&\textbf{Rule}&\textbf{support}&\textbf{confidence}&\textbf{lift}&\textbf{frequency}\\
        \midrule
        \multicolumn{6}{c}{\textbf{Simple where-questions}}\\
        \midrule
            1 & \{Q-6\} =\textgreater \{A-9\}     & 0.21    & 0.55       & 1.01 & 417   \\
            2 & \{Q-6\} =\textgreater \{A-8\}     & 0.20    & 0.54       & 1.24 & 404   \\
            3 & \{A-7\} =\textgreater \{A-9\}     & 0.16    & 0.56       & 1.73 & 307   \\
            4 & \{Q-6\} =\textgreater \{A-7\}     & 0.15    & 0.54       & 1.37 & 295   \\
            5 & \{A-7\} =\textgreater \{A-8\}     & 0.15    & 0.54       & 1.25 & 293   \\
        \midrule
        \multicolumn{6}{c}{\textbf{Detailed where-questions}}\\
        \midrule
            1 & \{Q-8\} =\textgreater \{A-7\}     & 0.65    & 0.80       & 1.08 & 1277  \\
            2 & \{Q-6\} =\textgreater \{Q-8\}     & 0.49    & 0.81       & 0.99 & 952   \\
            3 & \{Q-6\} =\textgreater \{A-7\}     & 0.48    & 0.80       & 1.07 & 940   \\
            4 & \{A-9\} =\textgreater \{Q-8\}     & 0.45    & 0.87       & 1.07 & 887   \\
            5 & \{A-7,Q-6\} =\textgreater \{Q-8\} & 0.42    & 0.88       & 1.07 & 823  \\
        \bottomrule
    \end{tabular}
    }
\end{table}

\begin{table}
\centering
\caption{\label{tab:prules}Extracted rules from prominence sequences}
\resizebox{0.75\textwidth}{!}{
    \begin{tabular}{llllll}
    \toprule
        \textbf{Rank}&\textbf{Rule}&\textbf{support}&\textbf{confidence}&\textbf{lift}&\textbf{frequency}\\
    \midrule
    \multicolumn{6}{c}{\textbf{Simple where-questions}}\\
    \midrule
        1 & \{A-4\} =\textgreater \{A-7\}     & 0.14    & 0.54       & 1.09 & 425   \\
        2 & \{A-5\} =\textgreater \{A-7\}     & 0.13    & 0.50       & 1.02 & 417   \\
        3 & \{Q-3\} =\textgreater \{A-7\}     & 0.12    & 0.52       & 1.05 & 382   \\
        4 & \{Q-6\} =\textgreater \{A-7\}     & 0.08    & 0.58       & 1.18 & 260   \\
        5 & \{Q-4\} =\textgreater \{A-7\}     & 0.08    & 0.53       & 1.07 & 250   \\
    \midrule
    \multicolumn{6}{c}{\textbf{Detailed where-questions}}\\
    \midrule
        1 & \{Q-6\} =\textgreater \{A-4\}     & 0.32    & 0.56       & 1.19 & 957   \\
        2 & \{Q-6\} =\textgreater \{A-7\}     & 0.30    & 0.51       & 1.06 & 884   \\
        3 & \{A-4\} =\textgreater \{A-7\}     & 0.25    & 0.54       & 1.11 & 742   \\
        4 & \{Q-3\} =\textgreater \{Q-6\}     & 0.24    & 0.54       & 0.93 & 695   \\
        5 & \{Q-3\} =\textgreater \{A-7\}     & 0.22    & 0.51       & 1.04 & 650  \\
    \bottomrule
    \end{tabular}
    }
\end{table}

Table~\ref{tab:trules} shows the dominant role of administrative divisions in the human-generated answers. Association rules extracted based on the scale (Table~\ref{tab:srules}) show the \emph{greater-than} and \emph{between} levels of the answers to SWQs and DWQs. The top five patterns of answers are mostly constructed with references to the highest level of prominence (\emph{A-7}). This shows the major impact of prominence in human answering behavior to where-questions -- i.e., people refer to prominent places in answering where-questions.

Tables~\ref{tab:trules}, \ref{tab:srules} and \ref{tab:prules} show that stronger association rules with higher support are extracted from DWQs in comparison to SWQs. The rules show strong associations between antecedent and consequent parts of the extracted rules with lift value greater than one. The results show that stronger rules with higher confidence and support are extracted using \textit{scale} in comparison to \textit{type} and \textit{prominence}.

The tables only present the extracted rules with highest frequency and support. These tables show how a small set of generic rules describes a large proportion of data in the MS MARCO dataset. Sorting rules by confidence or lift will change the order of the rules. For example, the maximum lift (equal to \textit{8.93}) in the extracted rules belongs to \{Q-6, Q-9\} =\textgreater \{A-8\} for detailed-where questions using scale. The frequency of this rule is \textit{43}, and it describes the relevant scale level (between minimum and maximum levels of the question) for detailed where-questions. The maximum confidence is 0.93 for detailed where-questions encoded by type. This association rule is \{Q-PPLA2, Q-ADM1\}	=\textgreater \{A-ADM2\} with a frequency \textit{109}. This rule describes that \textit{populated places} in detailed where-questions are mostly localized by referring to \textit{counties} they belong to.

\subsection{Predicting the Generic Form of Answers}
We test the predictability of the generic sequence of an answer given the generic sequence of the corresponding question. We investigate different prediction scenarios, including (1) the same generic class prediction (e.g., predicting type sequence of answers using type sequence of questions), and (2) prediction of one generic class using all generic classes (e.g., predicting type sequence of answers using type/scale/prominence sequences of questions, see Algorithm~\ref{alg:tsp}).

We assess the prediction accuracy of \textit{content} and \textit{content-and-style} of the answers (defined in Section 2.2.1). Referring to the introductory example, if type sequence of the answer is predicted as [river, city] then it is captured as a correct prediction for both content and content-and-style. The other permutation of this sequence (i.e., [city, river]) is considered as a correct prediction of content and incorrect prediction for content-and-style. Evidently, any other type sequence is an incorrect prediction for both content and content-and-style scenarios.

Each prediction scenario is applied over all questions, SWQs and DWQs to investigate the impacts of \textit{question types} on the prediction accuracy. Each scenario is tested using all six sequence prediction methods and is compared with the two baseline approaches (i.e., random generation and most frequent patterns). Only the best prediction performances among the six sequence prediction methods are presented. The \textit{best performance} is the maximum prediction accuracy achieved by one of the methods for a prediction scenario. We also test prediction accuracy when multiple predictions are allowed -- i.e., \textit{top-k predictions} for $k$ from one to five. In top-k predictions, $k$ unique sequences are predicted for each answer and if one of the sequences matches with the generic form of the answer then the prediction is successful.

Table~\ref{tab:tp_bests} shows the best performances in predicting type sequences of answers.The prediction accuracy based of TSP sequences is noticeably higher than that of predictions using only type sequences. This shows a complementary role of scale and prominence in predicting type sequence of the answers.

Contrasting DWQs and SWQs shows that extra details in DWQs are useful for prediction of the generic form of answers. In addition, we observe how subjectivity in style of answers and flexibility of language to convey information lead to noticeable less accuracy in prediction of content-and-style of answers in comparison to prediction of content. This observation is related to the flexibility of natural language, in which the same meaning can be presented in different ways. Finally, the number of predictions ($k$ in the table) shows that the accuracy dramatically increases in the case of multiple predictions. 

\begin{table}
\centering
\caption{\label{tab:tp_bests}Prediction accuracy for type sequences}
\resizebox{0.85\textwidth}{!}{
    \begin{tabular}{lcccc}
    \toprule
        \textbf{\#Predictions (k)} & \multicolumn{2}{c}{\textbf{Content}} & \multicolumn{2}{c}{\textbf{Content and Style}} \\ 
    \midrule
        & Type $\rightarrow$ Type & TSP $\rightarrow$ Type & Type $\rightarrow$ Type & TSP $\rightarrow$ Type \\ 
    \midrule
    \multicolumn{5}{c}{\textbf{All questions}} \\ 
    \midrule
        1 & 45.2 & 55.7 & 29.0 & 40.7 \\
        2 & 68.9 & 77.1 & 44.6 & 60.5 \\
        3 & 80.2 & 83.3 & 57.8 & 73.3 \\
        4 & 83.6 & 84.7 & 64.0 & 76.1 \\
        5 & 84.4 & 85.5 & 68.3 & 77.4 \\
    \midrule
    \multicolumn{5}{c}{\textbf{Simple where-questions}} \\ 
    \midrule
        1 & 39.5 & 47.5 & 14.2 & 27.4 \\
        2 & 60.8 & 69.4 & 32.7 & 48.5 \\
        3 & 73.2 & 75.8 & 48.2 & 63.4 \\
        4 & 77.2 & 77.5 & 58.1 & 66.2 \\
        5 & 78.5 & 78.2 & 63.1 & 67.0 \\
    \midrule
    \multicolumn{5}{c}{\textbf{Detailed where-questions}} \\ 
    \midrule
        1 & 59.1 & 67.3 & 47.1 & 59.6 \\
        2 & 80.4 & 88.7 & 61.3 & 76.3 \\
        3 & 84.0 & 91.2 & 65.9 & 84.4 \\
        4 & 88.0 & 91.3 & 73.6 & 86.4 \\
        5 & 88.5 & 92.1 & 75.6 & 87.1 \\
    \bottomrule
    \end{tabular}
    }
\end{table}

Tables~\ref{tab:sp_bests} and~\ref{tab:pp_bests} show that compared to type sequence prediction, the TSP sequences contribute less effectively in predicting the prominence and scale sequences -- i.e., only slightly improve the prediction accuracy. When considering multiple predictions, TSP sequences lead to worse results than prominence sequences or scale sequences alone. This can be explained by overfitting to specific patterns in the training dataset. Here, overfitting is observed because the schema of types is more than 20 times larger than the scale and prominence schemas. Hence, using type in prediction of scale or prominence leads to very detailed patterns that are not generalizable enough and decrease the prediction accuracy on unseen data. Finally, scale is the most predictable, and prominence is the least predictable generic class. Similar to the observations based on type prediction performances, DWQs are more predictable than SWQs based on scale and prominence.

\begin{table}
\centering
\caption{\label{tab:sp_bests}Prediction accuracy for scale sequences}
\resizebox{0.9\textwidth}{!}{
    \begin{tabular}{lllll}
    \toprule
        \textbf{\#Predictions (k)} & \multicolumn{2}{c}{\textbf{Content}} & \multicolumn{2}{c}{\textbf{Content and Style}} \\ 
    \midrule
        & Scale $\rightarrow$ Scale & TSP $\rightarrow$ Scale & Scale $\rightarrow$ Scale& TSP $\rightarrow$ Scale\\ 
    \midrule
    \multicolumn{5}{c}{\textbf{All questions}} \\ 
    \midrule
        1  & 55.0 & 56.7 & 38.2 & 42.2 \\
        2  & 79.4 & 79.2 & 61.0 & 62.8 \\
        3  & 91.6 & 86.1 & 79.0 & 76.0 \\
        4  & 96.3 & 88.7 & 92.0 & 81.9 \\
        5  & 98.0 & 89.3 & 96.0 & 83.5 \\
    \midrule
    \multicolumn{5}{c}{\textbf{Simple where-questions}} \\ 
    \midrule
        1 & 48.5 & 49.5 & 20.4 & 28.6 \\
        2 & 79.6 & 71.8 & 49.1 & 49.8 \\
        3 & 89.9 & 78.3 & 71.9 & 67.0 \\
        4 & 95.6 & 81.8 & 90.3 & 74.0 \\
        5 & 97.5 & 82.6 & 94.9 & 75.5\\
    \midrule
    \multicolumn{5}{c}{\textbf{Detailed where-questions}} \\ 
    \midrule
        1   & 69.6 & 68.2 & 59.8 & 60.6 \\
        2   & 88.4 & 89.6 & 78.4 & 77.3 \\
        3   & 95.8 & 93.3 & 88.6 & 87.1 \\
        4   & 97.5 & 95.2 & 94.8 & 92.1 \\
        5   & 98.6 & 95.2 & 97.0 & 92.7 \\
    \bottomrule
    \end{tabular}
    }
\end{table}

\begin{table}
\centering
\caption{\label{tab:pp_bests}Prediction accuracy of prominence sequences}
\resizebox{\textwidth}{!}{
    \begin{tabular}{lllll}
    \toprule
        \textbf{\#Predictions (k)} & \multicolumn{2}{c}{\textbf{Content}} & \multicolumn{2}{c}{\textbf{Content and Style}} \\ 
    \midrule
        & Prominence $\rightarrow$ Prominence & TSP $\rightarrow$ Prominence & Prominence $\rightarrow$ Prominence  & TSP $\rightarrow$ Prominence \\ 
    \midrule
    \multicolumn{5}{c}{\textbf{All questions}} \\ 
    \midrule
        1 & 50.8 & 53.0 & 19.9 & 30.7 \\
        2 & 74.1 & 73.4 & 39.2 & 49.1 \\
        3 & 85.0 & 81.9 & 61.6 & 66.4 \\
        4 & 92.1 & 86.7 & 79.2 & 77.1 \\
        5 & 96.1 & 88.6 & 89.2 & 81.8 \\
    \midrule
    \multicolumn{5}{c}{\textbf{Simple where-questions}} \\ 
    \midrule
        1 & 45.4 & 45.6 & 14.3 & 19.5 \\
        2 & 75.4 & 69.4 & 34.9 & 39.1 \\
        3 & 84.7 & 77.0 & 54.5 & 56.9 \\
        4 & 91.3 & 80.5 & 73.7 & 68.2 \\
        5 & 95.6 & 81.9 & 87.9 & 72.7 \\
    \midrule
    \multicolumn{5}{c}{\textbf{Detailed where-questions}} \\ 
    \midrule
        1 & 53.3 & 58.2 & 26.9 & 43.8 \\
        2 & 75.1 & 80.0 & 50.9 & 60.4 \\
        3 & 86.0 & 88.9 & 70.9 & 78.4 \\
        4 & 93.1 & 93.0 & 82.5 & 87.0 \\
        5 & 96.8 & 95.4 & 91.9 & 92.6 \\
    \bottomrule
    \end{tabular}
    }
\end{table}

Table~\ref{tab:seqPredBaselineComparision} shows the improvement of accuracy in best prediction performances compared to two baselines -- i.e., random generator, and most frequent pattern(s). The minimum improvement is +18.3\% in prediction of type sequences of answers using type sequences of questions in comparison to the most frequent pattern(s). This observation shows that strong patterns exist in the distributions of answers and consequently, the baseline method performs well in prediction of type sequences of answers. The strongest improvement is +61.6\% when comparing the best predictive performance of type sequences using type/scale/prominence sequences together, compared to the random baseline. This is because of the large number of distinct types in type schema that lead to false predictions for the random baseline. The accuracy improvements illustrate the strong relationship between the generic content of questions and generic content of their answers.

\begin{table}
\centering
\caption{\label{tab:seqPredBaselineComparision}Accuracy improvement using sequence prediction compared to the baselines}
\resizebox{0.8\textwidth}{!}{
    \begin{tabular}{lll}
    \toprule
        \textbf{Prediction  Scenario} & \textbf{Random} & \textbf{Most Frequent Pattern(s)} \\
    \midrule
        Type $\rightarrow$ Type & +48.9\% & +18.3\%\\
        Scale $\rightarrow$ Scale & +58.1\% & +27.6\%\\
        Prominence $\rightarrow$ Prominence & +39.2\% & +30.4\%\\
    \midrule
        TSP $\rightarrow$ Type & +61.6\% & +31.0\%\\
        TSP $\rightarrow$ Scale & +54.1\% & +23.6\%\\
        TSP $\rightarrow$ Prominence & +42.3\% & +33.5\%\\
    \midrule
        Overall & +50.7\% & +27.4\%\\
    \bottomrule
    \end{tabular}
    }
\end{table}

To compare the sequence prediction methods, we used the difference between the prediction accuracy of each method to the best performance achieved by all methods for each prediction scenario. Table~\ref{tab:seqMethodsRMSE} shows the root mean square error (RMSE) for each sequence prediction method. The RMSE shows how well-performed a method is in comparison to other methods. If the RMSE of a method is lower than others, the prediction accuracy of the method is higher than the others. The prediction scenarios in Table~\ref{tab:seqMethodsRMSE} are simplified groups of actual predictions. For example, prediction scenario of scale is related to predicting scale sequences of answers using (1) scale sequence of questions or (2) type/scale/prominence sequences of questions.

As shown in Table~\ref{tab:seqMethodsRMSE}, in all scenarios the \textbf{CPT} method is the \textit{best} performing method and \textbf{TDAG} performs \textit{worst} based on the RMSE values. The results suggest that \textbf{CPT} is the best method to construct predictive models to predict the generic form of answers.

\begin{table}
\centering
\caption{\label{tab:seqMethodsRMSE}RMSE of sequence prediction methods}
\resizebox{\textwidth}{!}{
	\begin{tabular}{cccccccc}
	\toprule
        \textbf{Prediction Scenario} & \textbf{LZ78} & \textbf{Mark1}   & \textbf{TDAG}  & \textbf{DG}    & \textbf{AKOM}  & \textbf{CPT}  & \textbf{CPT+}  \\
    \midrule
        Type & 7.4\% & 15.2\% & \cellcolor{red!25}\textbf{21.8\%} & 13.4\% & 17.3\% & \cellcolor{green!25}\textbf{7.1\%} & 12.9\% \\
        Scale & 9.8\% & 12.5\% & \cellcolor{red!25}\textbf{17.9\%} & 10.6\% & 14.3\% & \cellcolor{green!25}\textbf{5.7\%} & 11.8\% \\
        Prominence & 8.7\% & 13.3\% & \cellcolor{red!25}\textbf{19.2\%} & 9.9\%  & 15.2\% & \cellcolor{green!25}\textbf{4.9\%} & 11.5\% \\
    \midrule
        Content & 8.9\% & 15.5\% & \cellcolor{red!25}\textbf{22.7\%} & 9.1\%  & 17.4\% & \cellcolor{green!25}\textbf{1.9\%} & 10.2\% \\
        Content and Style & 8.6\% & 11.7\% & \cellcolor{red!25}\textbf{16.1\%} & 13.2\% & 13.7\% & \cellcolor{green!25}\textbf{8.2\%} & 13.8\% \\
    \bottomrule
    \end{tabular}
    }
\end{table}

\section{Demonstration: From Generic to Specific}

Translating generic encoding of answer to specific form (e.g., type sequence to toponym sequence) is the last phase in the proposed approach. Our approach to the generic-to-specific translation problem is grounded in the following assumption: \emph{places mentioned in the questions have relationships to places referred to in their answers, and these relations can be found in a knowledge base}. In addition, the specific form of questions and generic form of answers are available through encoding and prediction, respectively. Based on this assumption and the available information, the specific form of answer can be derived using a SPARQL query template (Query~\ref{lst:g_sparql}). While the \emph{structure} of a suitable knowledge base for this purpose has been studied before by \cite{Chen2018}, no such knowledge base is yet available with the definitions of type, scale and prominence as used in this study. Hence, the translation is only demonstrated here using the introductory example\footnote{More examples are provided in the supplementary material (Section 4)}.

We have used DBPedia and Geonames as sources to demonstrate how SPARQL queries can be used to find specific forms of answers. Considering the information stored in DBPedia and Geonames, this demonstration is limited to type sequences of the answers because the prominence and scale are not available in the place ontology of these knowledge bases. Even the type schema used in DBPedia is different from the Geonames' type schema, and consequently in the following example, mapping to the DBPedia type schema is done manually.

\begin{minipage}{0.9\linewidth}
    \begin{lstlisting}[captionpos=b, caption={SPARQL template}, label={lst:g_sparql},
    basicstyle=\scriptsize \ttfamily,frame=single]
    
    PREFIX [KNOWLEDGE BASE]
    
    SELECT distinct ?question ?answer 
    WHERE {
        VALUES ?question [SPECIFIC] .
        ?answer a [GENERIC] .
        {?question ?r ?answer} UNION {?answer ?r ?question} .
    }
    \end{lstlisting}
\end{minipage}

Referring to the introductory example, the where-question and its answer is modelled as follows:
\begin{itemize}
    \item specific representation (question): [Putney Bridge];
    \item TSP encoding (question): type sequence [BDG], scale sequence [4], prominence sequence [3];
    \item TSP encoding (answer): type sequence [STM, ADM2], scale sequence [6, 6], prominence sequence [6, 7];
    \item specific representation (answer): [River Thames, London]
\end{itemize}

The SPARQL queries for finding the specific forms of answers are presented in Queries \ref{lst:sparql3} and \ref{lst:sparql3_g} using DBPedia and Geonames ontologies. The results of these queries are shown in Table \ref{tab:eg3}. Using DBPedia, the generic forms are correctly translated into River Thames and London. However, the generic to specific translation using Geonames is partially successful. In Geonames, places are conceptualized as points and it supports only containment. This example shows that point-based conceptualization of places is not sufficient for generic to specific translation and more diverse support of spatial relationships can be useful to find the correct specific forms.

\begin{minipage}{0.9\linewidth}
\begin{lstlisting}[captionpos=b, caption={SPARQL query of the example (DBPedia)}, label={lst:sparql3},
  basicstyle=\scriptsize \ttfamily,frame=single]
PREFIX dbo: <http://dbpedia.org/ontology/>

SELECT distinct ?q1 ?a1 ?a2 WHERE {
  VALUES ?q1 {<http://dbpedia.org/resource/Putney_Bridge>} 

  ?a1 a dbo:PopulatedPlace .
  {?a1 ?r1 ?q1} UNION {?q1 ?r1 ?a1} .

  ?a2 a dbo:River .
  {?a2 ?r2 ?q1} UNION {?q1 ?r2 ?a2} .
}
\end{lstlisting}
\end{minipage}

\begin{minipage}{0.9\linewidth}
\begin{lstlisting}[captionpos=b, caption={SPARQL query of the example (Geonames)}, label={lst:sparql3_g},
  basicstyle=\scriptsize \ttfamily,frame=single]
PREFIX gn: <http://www.geonames.org/ontology#>

SELECT distinct ?q1 ?a1 ?a2 WHERE {
  VALUES ?q1 {<http://sws.geonames.org/6619925/>} 

  ?a1 gn:featureCode gn:A.ADM2 .
  {?a1 ?r1 ?q1} UNION {?q1 ?r1 ?a1} .

  ?a2 gn:featureCode gn:H.STM .
  {?a2 ?r2 ?q1} UNION {?q1 ?r2 ?a2} .
}
\end{lstlisting}
\end{minipage}

\begin{table}\centering
	\caption{\label{tab:eg3}SPARQL results to find specific form of the answer}
\begin{tabular}{lllll}\toprule
\textbf{Knowledge Base} & \textbf{Q1}           & \textbf{A1}       & \textbf{A2} \\
\midrule
DBPedia        & Putney Bridge       & \cellcolor{green!25}\textbf{London}  & \cellcolor{green!25}\textbf{River Thames} \\
\midrule
Geonames       & Putney Bridger & \cellcolor{green!25}\textbf{London}  & \cellcolor{red!25}\textbf{--} \\
\bottomrule
\end{tabular}
\end{table}

\section{Discussion}

The results of the proposed method shows how generic information can be used to characterize and imitate human answering behavior to generate templates for answering the questions. While the results are limited to the human-search engine interaction, the proposed methodology (specific-generic translation) is flexibly defined to be applicable to other QA scenarios as well.

We have used type, scale and prominence as generic classes to investigate MS~MARCO dataset. We have compared their potentials in describing human answering behavior and their performances in predicting the generic forms of the answers. As a result, two major observations are reported. 

First, while strong patterns for each generic class have been observed, we find that \textit{scale} is the most predictive class. This is because where-questions are a specific subset of spatial questions and scale directly captures inherent spatial aspect of places. Meanwhile, in the notions of type and prominence, other aspects of places contribute as well -- e.g., functional and physical aspects. In addition, scale is a generic class that captures hierarchical relationships between places, and previous studies show that these relationships are the basis for answering where-questions \citep{shanon1983answers}. Moreover, we have observed that type is performing better than prominence in both characterizing and predicting human-generated answers. This observation is highly influenced by the proxies used to capture type and prominence.

Second, when comparing SWQs and DWQs, our investigation shows that the generic templates to answer to DWQs, compared to SWQs, can be generated more accurately. We find stronger rules and patterns in the answers to DWQs than in answers to SWQs. This is because DWQs contain richer details which helps narrowing down the list of possible relevant answers. To illustrate this point, two examples are provided (1) \emph{Where in California is Beverly Hills?} and (2) \emph{Where is Beverly Hills?} In the first question, the list of possible relevant answers is narrowed down to \emph{Los Angeles County} because the inquirer already knows it is in \emph{California}. For the latter, respondents are free to subjectively guess the state of the inquirer's geographic knowledge and provide answers such as \emph{Los Angeles County}, \emph{California}, and \emph{United States}.

\textbf{Theoretical limitations:} The instruction for specific-generic approach is devised in a flexible manner to be usable in different GeoQA scenarios. However, utilizing the approach needs a careful design (e.g., selecting appropriate list of generic classes) to fit for a particular scenario. The proposed TSP encoding is limited to the QA scenario of general Web search and may not be suitable for other QA scenarios such as human interaction with autonomous cars. In short, the theoretical limitations of this study are:

\begin{enumerate}
    \item The generic-to-specific translation approach is only focused on where-questions, and other types of geographic questions are neglected.
    
    \item The proposed approach is focused only on the questions and their relationship with the answers, when no other contextual information about inquirers is available.
    
    \item The approach is designed with an exclusive focus on toponyms while qualitative spatial relationships have an important role in answering where questions.
    
    \item The additional impacts of qualitative spatial relationships (e.g., \emph{in southern part of}) as modifiers of scale are neglected in the TSP encoding.
\end{enumerate}

\textbf{Results limitations:} There are some limitations to the implementation presented in this study:
\begin{enumerate}
   \item The biases of the MS MARCO dataset directly influence our results. The data are extracted from the Microsoft Bing search engine, and hence the results are necessarily biased to the questions asked by users of this search engine. In addition, the sampling approach used when extracting MS MARCO questions from the MS Bing query logs may have a direct and unquantifiable impact on the generality of the results.
   
   \item The results are influenced by the geographic biases and incompleteness of data in Geonames and OSM Nominatim. The bias and incompleteness of gazetteers are well-documented by \cite{ACHESON2017309}.
   
   \item The bias in the proxies that have been used to capture the TSP encoding also have an impact on the results.
\end{enumerate}

Despite these limitations, the identified patterns align well with everyday experience and provide a grounding for answering where-questions.

\section{Conclusions}
Generating responses with a similar quality to human-generated answers is a challenge to current search engines and QA systems. In particular, where-questions are hard to answer because the responses can sometimes be either vague or obvious to the inquirers. To avoid generating ambiguous or obvious responses or retrieving unnecessary information as a part of the answers, a proper set of anchor places must be identified to localize the place in question. The assumption that answers to where-questions can be found completely, without any further modification, inside a textual document or as a node or its properties in a knowledge base may not hold in general. Consequently, we introduced here an approach to generate templates to answer where-questions based on relevant pieces of information.

The approach is based on the automatic extraction of patterns of generic geographic forms from human-generated QA. These captured in predictive models and are used to generate templates of answers similar human-generated responses. Three generic classes (i.e., type, scale and prominence) are used to investigate the properties of the anchor places in human-generated answers. We have used questions and answers from MS MARCO v2.1, an extensive dataset constructed from questions submitted to a general-purpose search engine. 

Using distribution analysis and rule mining techniques, we have identified the characteristics and recurrent patterns in the questions and their answers (Hypotheses 1 and 2). We have then applied sequence prediction methods to generate the generic forms for answers based on the generic forms of the corresponding questions (Hypothesis 3). We have also briefly sketched an approach how such generic forms may help with the generation of the appropriate answers, based on the information available in the knowledge bases. 

The results show that the prediction of answer structures based on \textit{scale} is more precise, compared to predictions relying on \textit{type} and \textit{prominence}. The rules extracted based on scale have higher support and confidence than the rules extracted from type or prominence. We also observe how the type of questions (i.e., SWQs vs. DWQs) influence the strength of the extracted rules and lead to noticeable differences in prediction performances. Finally, we compared different sequence prediction methods and find that CPT \citep{gueniche2013compact} is the best performing approach in all scenarios. However, the results of this study are limited to human interaction with a general purpose search engine. Consequently, an important future direction of this research is to investigate other corpora of QA related to different scenarios -- e.g., human-human dialogue.

We have also observed that the neglect of qualitative spatial relationships in our encoding and prediction mechanism may present a a major theoretical shortcoming of the proposed specific-generic translation. Consequently, developing a more sophisticated encoding is necessary to extract a deeper understanding of the human answering behavior of where-questions. 

Developing an automatic approach to decode generic forms of answers into specific representations (i.e., toponyms) is a necessary step to complete the process of the specific-generic translation approach. Available information in documents or knowledge bases can be used to derive the specific representations. Another important future direction is to investigate how the proposed approach can be combined with current personalization methods, in order to adapt answers to specific inquirers and their context. Finally, investigation of other types of where-questions (i.e., where-questions with generic references) and their human-generated answers using specific-generic translation remains as a future work.

\section{Data and Codes Availability Statement}
This study makes use of a third-party data source, MS MARCO v2.1 \citep{Nguyen2016ms}. The dataset is freely available under a proprietary agreement for non-commercial use\footnote{\url{http://www.msmarco.org/dataset.aspx}}. The computational workflow of this publication is implemented in Java and R. The implementation is available under the MIT License\footnote{\url{https://opensource.org/licenses/MIT}} and accessible in GitHub: \url{https://github.com/hamzeiehsan/Template-for-answering-where-questions}.

\section*{Acknowledgments}
The support by the Australian Research Council grant DP170100109 is acknowledged. We also thank the anonymous reviewers for their helpful comments that improve the quality of this paper.

\bibliographystyle{apacite}
\bibliography{main}

\end{document}



\title{Supplementary material: Templates of generic geographic information for answering where-questions}

\author{
\name{Ehsan Hamzei, Stephan Winter and Martin Tomko\textsuperscript{a}}\\

\affil{\textsuperscript{a}The University of Melbourne}
}

\maketitle

This is a supplementary material for ``Templates of generic geographic information for answering where-questions''. In this document, first details of the implementation are presented. Next,
 distribution and frequent patterns of type, scale and prominence in the questions and their answers are presented and discussed. Finally, four examples are provided to demonstrate deriving the specific form of the answers (toponyms) from their generic forms (e.g., type-sequence).

\section{Implementation}
\subsection{Scale Encoding}
In scale encoding, the scale values of OpenStreetMap (OSM) are mapped into the proposed scale schema. The mapping is shown in Table \ref{tab:scaleMapping}.

\begin{table}
\centering
\caption{\label{tab:scaleMapping}Scale mapping from OSM values to proposed schema}
\begin{tabular}{llllll}\midrule
Scale level (proposed schema) & OSM scale levels & Description\\
\midrule
1&	value $\geq$ 27 & buildings and houses\\
2& 22 $\leq$ value $<$ 27 & airports, roads and streets\\
3& 18 $\leq$ value $<$ 22 & suburbs and districts\\
4& 16 $\leq$ value $<$ 18 & cities, islands and villages\\
5& 12 $\leq$ value $<$ 16 & counties\\
6& 8 $\leq$ value $<$ 12 & states and regions\\
7& 4 $\leq$ value $<$ 8 & countries\\
8& value $<$ 4 & oceans, seas and continents\\
\bottomrule
\end{tabular}
\end{table}

\subsection{Prominence Encoding}
\begin{figure}
\centering
\includegraphics[width=0.7\textwidth]{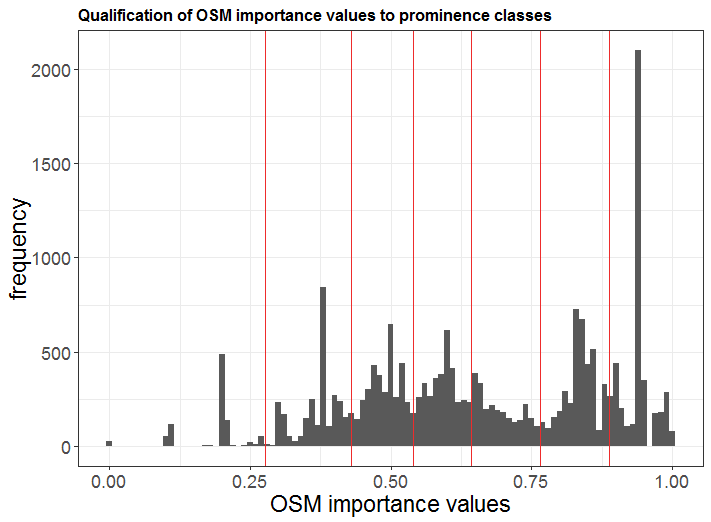}
\caption{Histogram of importance values and boundaries of prominence levels.}
\label{fig:prominenceLevels}
\end{figure}

To derive prominence levels from OSM importance values, we have used `natural breaks' classification. First, all the importance values are collected, then based on the histogram the natural break method calculates the boundaries of each prominence level. Figure \ref{fig:prominenceLevels} shows the histogram and the derived boundaries for the prominence levels.

\section{Comparing Simple-Where Questions and Detailed Where-Questions}
\begin{figure}
    \centering
    \includegraphics[width=\textwidth]{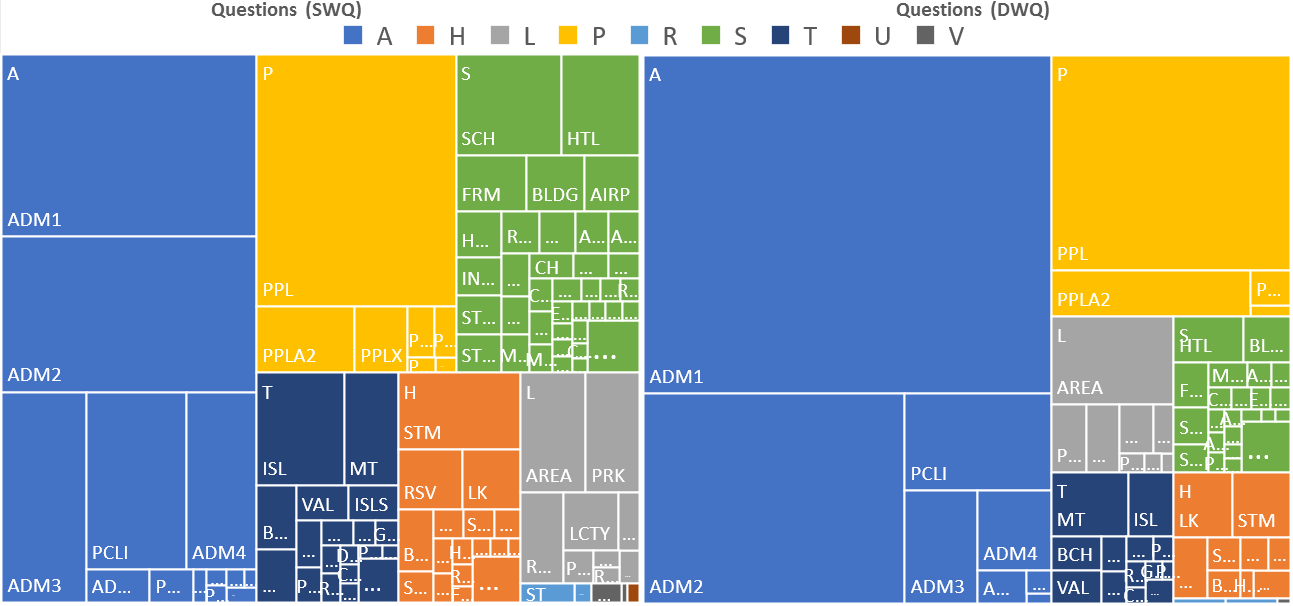}
    \caption{Distribution of types in the SWQs versus DWQs.}
    \label{fig:tswqvsdwq}
\end{figure}

Figure~\ref{fig:tswqvsdwq} shows the distributions of place types in SWQs and DWQs. In DWQs, administrative divisions are almost twice as frequent as in SWQs. The extra detail (additional toponyms) included in the content of DWQs dominantly refers to administrative divisions. For example, \emph{Where in California is Disneyland?} can be viewed as a SWQ, \emph{Where is Disneyland?}, with extra details that shows what the inquirer already knows about its coarse location (\emph{in California}). Hence, the differences between the two distributions illustrate where the additional detail is included.

\begin{figure}
    \centering
    \includegraphics[width=\textwidth]{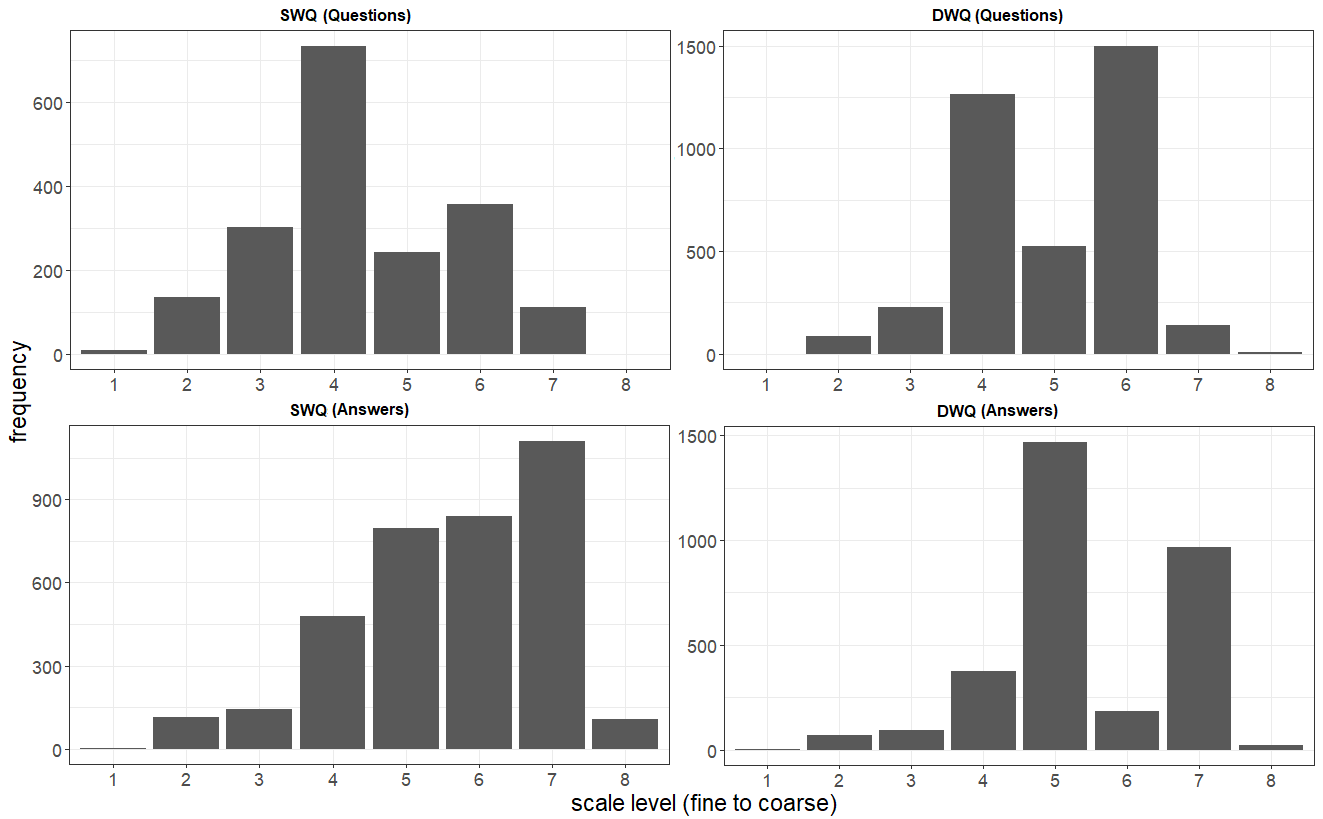}
    \caption{Distribution of the levels of scale in the SWQs versus DWQs.}
    \label{fig:sswqvsdwq}
\end{figure}

Answers to SWQs (Figure~\ref{fig:sswqvsdwq}) have a linearly incremental distribution (except for the coarsest level of scale). The figure shows the answers are generated with toponyms that belong to \textit{coarser levels} of scale compared to places mentioned in the SWQs. The distribution of scale in the answers to DWQs shows a dominance of references to places of scale \textit{between} the two mean values of the questions' bi-modal distribution. Thus, people tend to generate responses at a level of scale greater than that of the place that is asked for, and lower than the scale of the places mentioned as additional details. This is because places at a same or coarser level of scale than the places in questions would lead to obvious, and hence irrelevant, answers. The clear distinction of the difference in answering SWQs and DWQs are shown in Table~\ref{tab:sswqvsdwq}, enabling the comparison of the levels of scale of places identified in the questions and their answers.

\begin{table}
    \centering
    \caption{\label{tab:sswqvsdwq}Comparison of levels of scale in the answers compared to their questions}
    \resizebox{0.9\textwidth}{!}{
    \begin{tabular}{llll}
    \toprule
        \textbf{Level of scale in answers}  & \textbf{Lower than} &\textbf{ Equal (Between)} & \textbf{Greater than} \\
    \midrule
    \multicolumn{4}{c}{\textbf{SWQ}}\\
    \midrule
        each value   & 17.0\%    &13.1\%  & \textbf{69.9\%}\\
        min value    & 25.8\%    & 16.0\%   & \textbf{58.2\%}\\
        median value & 14.5\%  & 14.9\%   & \textbf{70.6\%}\\
        max value    & 10.6\%  & 5.9\%  & \textbf{83.5\%}\\
    \midrule
    \multicolumn{4}{c}{\textbf{DWQ}}\\
    \midrule
        each value   & 7.2\% & \textbf{59.6\%} & 33.2\%\\
        min value    & 10.5\% & \textbf{77.7\%} & 11.8\%\\
        median value & 6.7\% & \textbf{77.3\%} & 16.0\%\\
        max value    &  3.9\% & 40.8\% & \textbf{55.3\%}\\
    \bottomrule
    \end{tabular}
    }
\end{table}

\begin{figure}
    \centering
    \includegraphics[width=\textwidth]{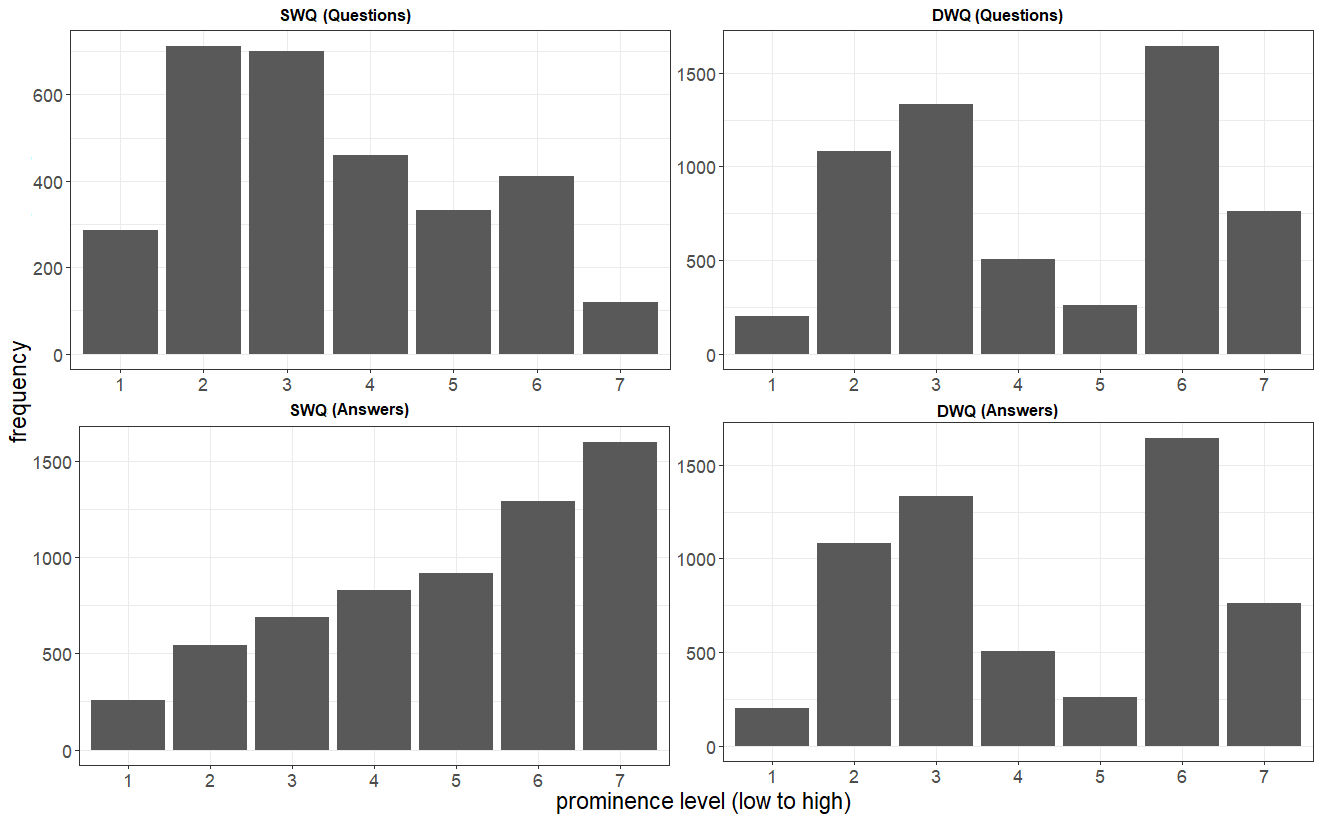}
    \caption{Distribution of prominence levels in the SWQs versus DWQs}
    \label{fig:pswqvsdwq}
\end{figure}

Figure~\ref{fig:pswqvsdwq} shows the distribution of prominence in SWQs, DWQs, and their answers. We observe that people ask about less-known places and answer referring to well-known ones. In DWQs, both less-known and well-known places are frequently observed. The well-known places in DWQs are the details presented in the content of the questions. In DWQs, inquirer's state of knowledge can be estimated better, because the question contains what they already know and what they want to find out. Similar to the pattern observed based on scale, prominence levels of places in the answers to DWQs are mostly between the levels of prominence of the toponyms mentioned in their corresponding questions (Table~\ref{tab:pswqvsdwq}).

\begin{table}
    \centering
    \caption{\label{tab:pswqvsdwq}Comparison of prominence levels in the answers compared to their questions}
    \resizebox{0.9\textwidth}{!}{
        \begin{tabular}{llll}
        \toprule
            \textbf{Prominence level in answers}  & \textbf{Lower than} & \textbf{Equal (Between)} & \textbf{Greater than} \\
        \midrule
        \multicolumn{4}{c}{\textbf{SWQ}}\\
        \midrule
            each value& 19.8\% & 12.5\% & \textbf{67.7\%}\\
            min value& 30.7\% & 14.7\% &\textbf{54.6\%}\\
            median value& 19.4\% & 9.5\% &\textbf{71.1\%}\\
            max value& 9.7\%& 7.9\% &\textbf{82.4\%}\\
        \midrule
        \multicolumn{4}{c}{\textbf{DWQ}}\\
        \midrule
            each value& 8.7\% & \textbf{68.4\%} & 22.9\% \\
            min value& 13.0\% & \textbf{77.9\%} & 9.1\% \\
            median value& 6.8\% & \textbf{79.7\%} & 13.5\% \\
            max value& 3.5\% & \textbf{55.6\%} & 40.9\%\\
        \bottomrule
    \end{tabular}
    }
\end{table}

\section{Sequence Distributions and Frequent Patterns}
\subsection{Sequence Distributions}
\begin{figure}
\centering
\includegraphics[width=\textwidth]{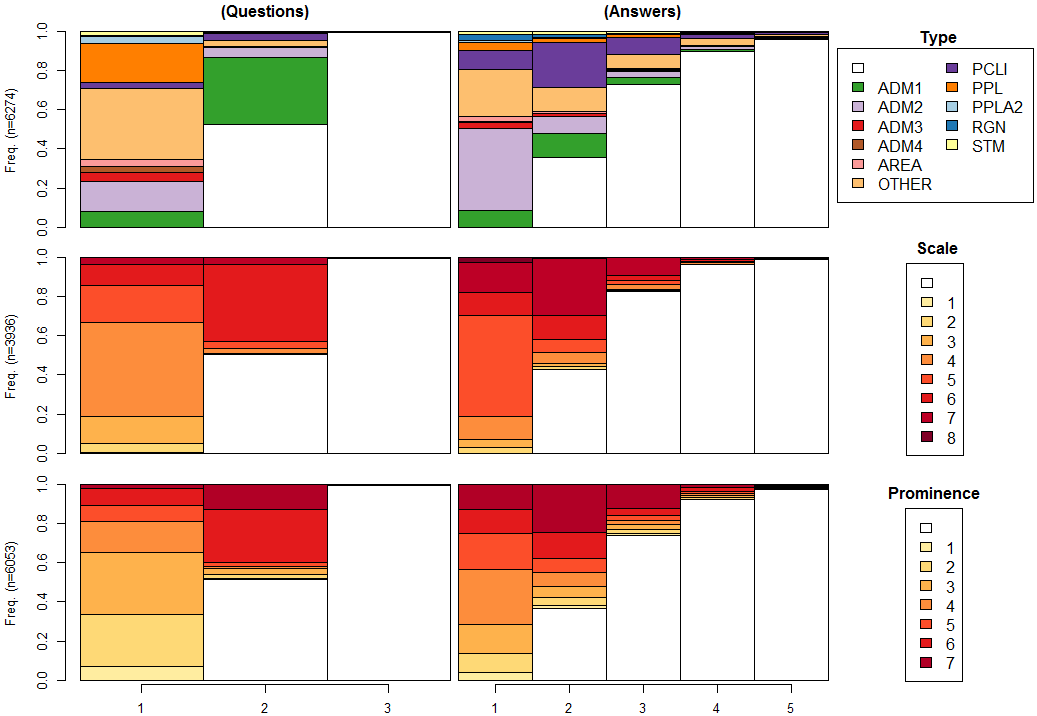}
\caption{Sequence distributions of the questions and answers}
\label{fig:d_tsp_qvsa}
\end{figure}

Sequence distributions capture not only the content of sequences, but
the way that the content is generated in the sequence (i.e., \emph{style}). The length of sequences can be further investigated using sequence distributions. Figure~\ref{fig:d_tsp_qvsa} shows the sequence distributions of type, scale and prominence in the questions and answers. Here, we visualize only the first five positions of answer-sequences. While this long-tailed distribution contains answers up to a length of 13, the vast majority (95.84\%) of answers have less than five toponyms. 
Most of the questions have only one or two toponyms, and the answers are mostly short as well (usually with less than three toponyms in their content). In Figure~\ref{fig:d_tsp_qvsa}, only the top ten types are visualized, with the rest grouped as \emph{OTHER}.

Based on the distributions of type, scale and prominence in the questions, the details in DWQs are likely to be found as the second toponym. In Figure~\ref{fig:d_tsp_qvsa}, one can observe that these details usually belong to the type of ADM1\footnote{The codes for place types are described in Section~\ref{appendix:a}.} (first-order administrative divisions -- i.e., states) or PCLI (independent political entities -- i.e., countries). Similar patterns can be observed using scale and prominence -- i.e., the second elements in questions belong to well-known places which are in coarse levels of scale. 

Figure~\ref{fig:d_tsp_qvsa} shows the strong role of administrative places in the human-generated answers in this corpus. The answers are generated with mid-levels to coarse-levels of scale and presented starting with mid-levels followed by coarse levels in most cases. Similar patterns can be observed in terms of prominence; however, places in lowest and highest prominence levels can be found in the answers. When comparing the questions and answers based on type, we observe that while the questions ask about a diverse range of place types, the answers are mostly generated using the top ten types (i.e., the less frequent types labelled as \emph{OTHER} are almost twice as frequent in the questions compared to answers).

\subsection{Frequent Patterns}
\begin{figure}
\centering
\includegraphics[width=\textwidth]{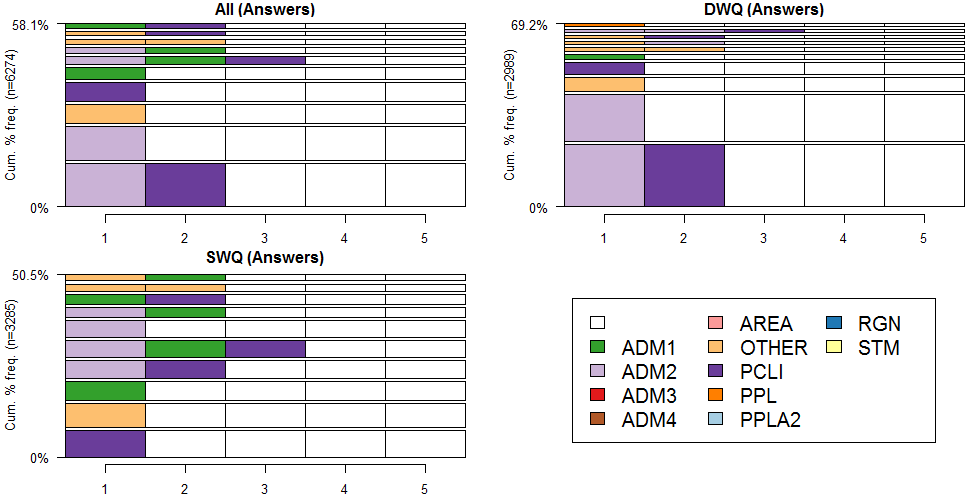}
\caption{Top ten frequent pattern in type-sequences of the answers}
\label{fig:t_sp}
\end{figure}

\begin{figure}
\centering
\includegraphics[width=\textwidth]{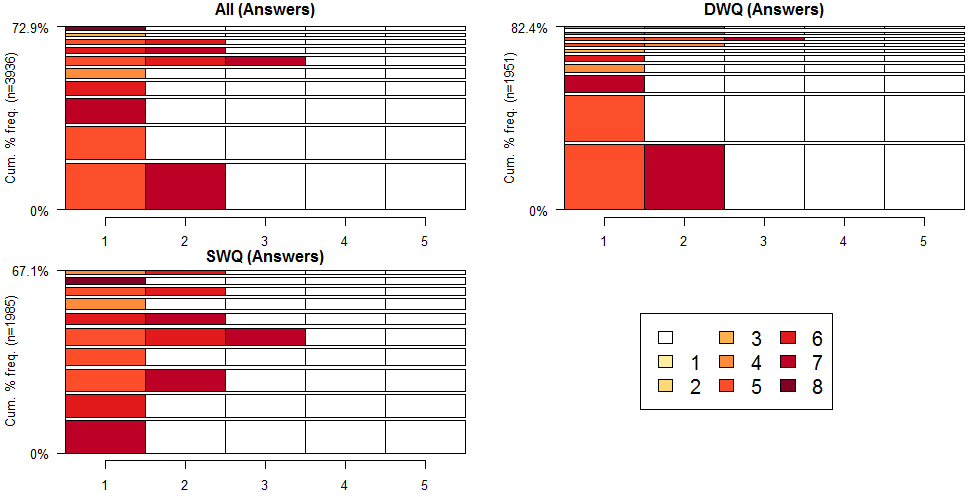}
\caption{Top ten frequent pattern in scale-sequences of the answers}
\label{fig:s_sp}
\end{figure}

\begin{figure}
\centering
\includegraphics[width=\textwidth]{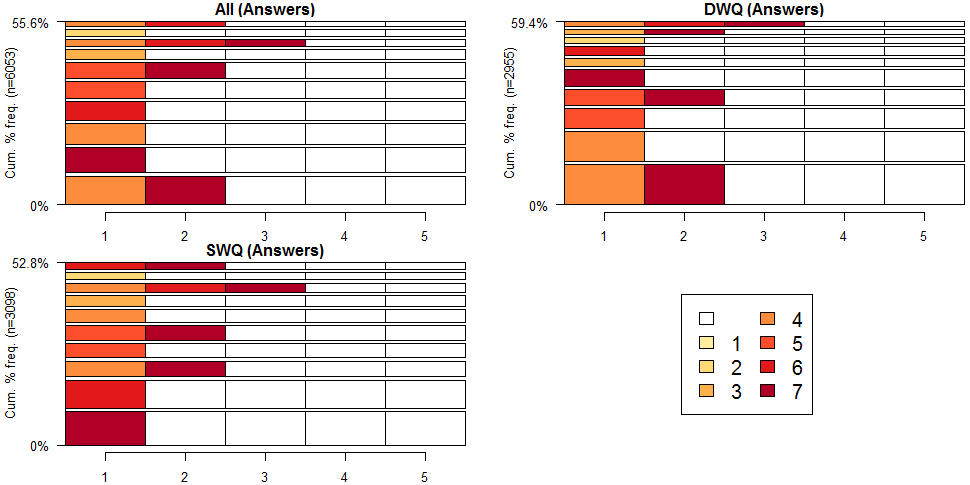}
\caption{Top ten frequent pattern in prominence-sequences of the answers}
\label{fig:p_sp}
\end{figure}

Frequent patterns describe the generalized patterns in human answering behavior. Stronger patterns of a generic class (e.g., scale) show the usefulness of the generic information to describe the human-generated answers. Figures~\ref{fig:t_sp}, \ref{fig:s_sp} and \ref{fig:p_sp} illustrate the top ten patterns based on type, scale and prominence, respectively. These patterns are shown for all answers, answers to SWQs, and answers to DWQs.

Figure~\ref{fig:t_sp} shows that the top ten patterns in the answers are mostly generated with first and second orders administrative divisions and independent political entities. In the generic class, \emph{type}, 
almost 60\% of all answers are described through only ten sequence patterns. Ten patterns also cover 70\% and 50\% of the answers to DWQs and SWQs, respectively. The differences shows that answers to DWQs are more describable based on place type than the responses to SWQs. This is because in DWQs, some details are provided and people can therefore infer the inquirers' information needs better. In SWQs, due to lack of context leads to more ambiguity and likely requires more subjective judgments in responses.

Figure~\ref{fig:s_sp} illustrates the top ten patterns based on scale. The coarse levels of scale constitute most of the frequent patterns. Answers covered by these patterns show that scale is an important generic information characterizing well human-generated answers. 
The comparison of the answers to DWQs and SWQs shows similar results to types. More than 80\% of the answers to DWQs can be described using ten patterns based on scale. The style of these patterns shows that the answers are hierarchically presented (starting from finer levels followed by coarser levels of scale).

Figure~\ref{fig:p_sp} shows the ten most frequent patterns based on prominence levels. Most of the patterns are constructed with high levels of prominence. Similar to scale, a strong pattern in the style of answers can be observed -- i.e., starting with less-known places followed by well-known ones. The patterns based on prominence have, however, less support in the data compared to the patterns derived from scale or type sequences. This observation can be explained by prominence being more tightly related to the specific context of the questions, compared to scale or type. For example, finding a highly prominent place reference may not be always possible
as prominent places are not uniformly distributed in the world.

\section{Demonstration Examples}
In this section, four examples are provided to demonstrate how generic form of answers (e.g., type-sequence) can be translated into their specific form (toponym-sequence).

\textbf{Example 1:} \emph{Where is Nagasaki?} \emph{In Japan}. The specific form and TSP encoding of this question-answer pair are shown below:
\begin{itemize}
    \item specific representation (question): [Nagasaki];
    \item TSP encoding (question): type-sequence [ADM1], scale-sequence [6], prominence-sequence [4];
    \item TSP encoding (answer): type-sequence [PCLI], scale-sequence [7], prominence-sequence [7];
    \item specific representation (answer): [Japan]
\end{itemize}

Queries \ref{lst:sparql1} and \ref{lst:sparql1_g} show the SPARQL queries to derive the specific form of the answer using the information available in DBPedia and Geonames, respectively. The result of both queries is the unified resource identifier (URI) of Japan which is the correct specific representation of the answer. The query results are shown in Table \ref{tab:eg1}.

\begin{minipage}{0.9\linewidth}
\begin{lstlisting}[captionpos=b, caption={SPARQL query of Example 1 (DBPedia)}, label={lst:sparql1},
  basicstyle=\scriptsize \ttfamily,frame=single]
PREFIX dbo: <http://dbpedia.org/ontology/>

SELECT distinct ?q1 ?a1 WHERE {
  VALUES ?q1 {<http://dbpedia.org/resource/Nagasaki>}  
 
  ?a1 a dbo:Country .
  ?q1 ?r ?a1
}

\end{lstlisting}
\end{minipage}

\begin{minipage}{0.9\linewidth}
\begin{lstlisting}[captionpos=b, caption={SPARQL query of Example 1 (Geonames)}, label={lst:sparql1_g},
  basicstyle=\scriptsize \ttfamily,frame=single]
PREFIX gn: <http://www.geonames.org/ontology#>

SELECT distinct ?q1 ?a1 WHERE  {
  VALUES ?q1 {<http://sws.geonames.org/1856156/>} . 
   
  ?a1 gn:featureCode gn:A.PCLI .
  ?q1 ?r ?a1 
}

\end{lstlisting}
\end{minipage}

\begin{table}\centering
	\caption{\label{tab:eg1}SPARQL results to Example 1}
\begin{tabular}{lll}\toprule
\textbf{Knowledge Base} & \textbf{Q1}       & \textbf{A1}    \\
\midrule
DBPedia        & Nagasaki & \cellcolor{green!25}\textbf{Japan} \\
\midrule
Geonames       & Nagasaki & \cellcolor{green!25}\textbf{Japan} \\
\bottomrule
\end{tabular}
\end{table}

\textbf{Example 2:} \emph{Where in Illinois is Cahokia?} \emph{In St. Clair County, Illinois, United States}:
\begin{itemize}
    \item specific representation (question): [Illinois, Cahokia];

    \item TSP encoding (question): type-sequence [ADM1, PPL], scale-sequence [6, 4], prominence-sequence [6, 3];
    
    \item TSP encoding (answer): type-sequence [ADM2, PCLI], scale-sequence [5, 7], prominence-sequence [3, 7];
    
    \item specific representation (answer): [St. Clair County, United States]
\end{itemize}

Based on the SPARQL template, two queries (Queries~\ref{lst:sparql2} and~\ref{lst:sparql2_g}) are used to translate the generic form of the answer to a specific form using DBPedia and Geonames. The results of the queries are shown in Table~\ref{tab:eg2}. Using DBPedia, the correct answer is among one of the three retrieved responses \{[Collinsville, United States], [Illinois, United States], [St. Clair County, United States]\}. The second response in the results can be easily filtered due to repetitive content (i.e., Illinois) considering the content of the question. However, the first one which is not mentioned in the human-generated answer cannot be avoided using only the type-sequence of the answer. Consequently, by considering the predicted scale-sequence of the answer, the correct response [St. Clair County, United State] can be derived (Table~\ref{tab:eg2}).

\begin{minipage}{0.9\linewidth}
    \begin{lstlisting}[captionpos=b, caption={SPARQL query of Example 2 (DBPedia)}, label={lst:sparql2},
    basicstyle=\scriptsize \ttfamily,frame=single]

    PREFIX dbo: <http://dbpedia.org/ontology/>

    SELECT distinct ?q1 ?q2 ?a1 ?a2 
    WHERE {
        VALUES ?q1 {<http://dbpedia.org/resource/Cahokia>}  
        VALUES ?q2 {<http://dbpedia.org/resource/Illinois>}  
        
        ?a1 a dbo:PopulatedPlace .
        ?q1 ?r ?a1 . 
        ?a1 ?r2 ?q2 .

        ?a2 a dbo:Country . 
        ?q2 ?r3 ?a2 .
    }
    \end{lstlisting}
\end{minipage}

\begin{minipage}{0.9\linewidth}
    \begin{lstlisting}[captionpos=b, caption={SPARQL query of Example 2 (Geonames)}, label={lst:sparql2_g},
    basicstyle=\scriptsize \ttfamily,frame=single]

    PREFIX gn: <http://www.geonames.org/ontology#>

    SELECT distinct ?q1 ?q2 ?a1 ?a2 
    WHERE  {
        VALUES ?q1 {<http://sws.geonames.org/4234969/>} . 
        VALUES ?q2 {<http://sws.geonames.org/4896861/>} . 
        
        ?a1 gn:featureCode gn:A.ADM2 .
        ?q1 ?r ?a1 . 
        ?a1 ?r2 ?q2 .
   
        ?a2 gn:featureCode gn:A.PCLI .
        ?q2 ?r3 ?a2 .
    }
    \end{lstlisting}
\end{minipage}

\begin{table}\centering
\caption{\label{tab:eg2}SPARQL results to Example 2}
\resizebox{0.7\textwidth}{!}{
    \begin{tabular}{lllll}
    \toprule
        \textbf{Knowledge Base} & \textbf{Q1}      & \textbf{Q2}       & \textbf{A1}               & \textbf{A2}           \\
    \midrule
        DBPedia        & Cahokia & Illinois & Collinsville     & \cellcolor{green!25}\textbf{United States} \\
               &         &          & Illinois         & \cellcolor{green!25}\textbf{United States} \\
               &         &          & \cellcolor{green!25}\textbf{St. Clair County} & \cellcolor{green!25}\textbf{United States} \\\midrule
        Geonames       & Cahokia & Illinois & \cellcolor{green!25}\textbf{St. Clair County} & \cellcolor{green!25}\textbf{United States}\\
    \bottomrule
    \end{tabular}
    }
\end{table}

\textbf{Example 3:} \emph{Where is the Danube River, Europe?} \emph{It originates in Germany's Black Forest, and flows in a southeasterly direction through central and eastern Europe to the Black Sea}:
\begin{itemize}
    \item specific representation (question): [Danube River, Europe];
    \item TSP encoding (question): type-sequence [STM, CONT], scale-sequence [4, 8], prominence-sequence [6, 7];
    \item TSP encoding (answer): type-sequence [PCLI, MTS, SEA], scale-sequence [7, 3, 7], prominence-sequence [7, 4, 5];
    \item specific representation (answer): [Germany, Black Forest, Black Sea]
\end{itemize}

The SPARQL queries for finding the specific forms of answers of Example 2 are presented in Queries \ref{lst:sparql3} and \ref{lst:sparql3_g}. The results of these queries are shown in Table \ref{tab:eg3}. Using DBPedia, the results for the mountain range and the sea that are related to the Danube are the Black Forest and the Black Sea, while the country is not unique and ten countries (including the right one, Germany, based on the human-generated answer) are found in relation to the river. Here, we have only used the type-sequence of the answer, and using prominence we could limit the country lists to Germany (the only country in Level 7 of prominence in the results). Interestingly, the results using Geonames are incorrect. In Geonames, the river is stored as a coordinate point, which belongs to one country. Moreover, the relationships between Black Forest (\emph{originates}) and Black Sea (\emph{flows to}) to the river are not stored in Geonames due to its limited list of supported spatial relations -- i.e., Geonames supports only containment.

\begin{minipage}{0.9\linewidth}
\begin{lstlisting}[captionpos=b, caption={SPARQL query of Example 3 (DBPedia)}, label={lst:sparql3},
  basicstyle=\scriptsize \ttfamily,frame=single]
PREFIX dbo: <http://dbpedia.org/ontology/>

SELECT distinct ?q1 ?a1 ?a2 ?a3 WHERE {
  VALUES ?q1 {<http://dbpedia.org/resource/Danube>} 

  ?a1 a dbo:Country .
  ?q1 ?r1 ?a1 .

  ?a2 a dbo:MountainRange .
  ?q1 ?r2 ?a2 .

  ?a3 a dbo:Sea .
  ?a3 ?r3 ?q1
}
\end{lstlisting}
\end{minipage}

\begin{minipage}{0.9\linewidth}
\begin{lstlisting}[captionpos=b, caption={SPARQL query of Example 3 (Geonames)}, label={lst:sparql3_g},
  basicstyle=\scriptsize \ttfamily,frame=single]
PREFIX gn: <http://www.geonames.org/ontology#>

SELECT distinct ?q1 ?a1 ?a2 ?a3 WHERE {
  VALUES ?q1 {<http://sws.geonames.org/791630/>} 

  ?a1 gn:featureCode gn:A.PCLI .
  ?q1 ?r1 ?a1 .

  ?a2 gn:featureCode gn:T.MTS .
  ?q1 ?r2 ?a2 .

  ?a3 gn:featureCode gn:H.SEA .
  ?q1 ?r3 ?a3 
}
\end{lstlisting}
\end{minipage}

\begin{table}\centering
	\caption{\label{tab:eg3}SPARQL results to Example 3}
\begin{tabular}{lllll}\toprule
\textbf{Knowledge Base} & \textbf{Q1}           & \textbf{A1}       & \textbf{A2}           & \textbf{A3}        \\
\midrule
DBPedia        & Danube       & Moldova  & \cellcolor{green!25}\textbf{Black Forest} & \cellcolor{green!25}\textbf{Black Sea} \\
              &              & Ukraine  & \cellcolor{green!25}\textbf{Black Forest} & \cellcolor{green!25}\textbf{Black Sea} \\
              &              & \cellcolor{green!25}\textbf{Germany}  & \cellcolor{green!25}\textbf{Black Forest} & \cellcolor{green!25}\textbf{Black Sea} \\
              &              & Romania  & \cellcolor{green!25}\textbf{Black Forest} & \cellcolor{green!25}\textbf{Black Sea} \\
              &              & Croatia  & \cellcolor{green!25}\textbf{Black Forest} & \cellcolor{green!25}\textbf{Black Sea} \\
              &              & Serbia   & \cellcolor{green!25}\textbf{Black Forest} & \cellcolor{green!25}\textbf{Black Sea} \\
              &              & Austria  & \cellcolor{green!25}\textbf{Black Forest} & \cellcolor{green!25}\textbf{Black Sea} \\
              &              & Bulgaria & \cellcolor{green!25}\textbf{Black Forest} & \cellcolor{green!25}\textbf{Black Sea} \\
              &              & Hungary  & \cellcolor{green!25}\textbf{Black Forest} & \cellcolor{green!25}\textbf{Black Sea} \\
              &              & Slovakia & \cellcolor{green!25}\textbf{Black Forest} & \cellcolor{green!25}\textbf{Black Sea} \\
\midrule
Geonames       & Danube River & Romania  & \cellcolor{red!25}\textbf{--}   & \cellcolor{red!25}\textbf{--} \\
\bottomrule
\end{tabular}
\end{table}

\textbf{Example 4:} \emph{Where is Golden Gate Bridge?} \emph{It is located between San Francisco and Marin County, in the U.S. state of California.}:
\begin{itemize}
    \item specific representation (question): [Golden Gate Bridge];

    \item TSP encoding (question): type-sequence [BDG], scale-sequence [5], prominence-sequence [4];
    
    \item TSP encoding (answer): type-sequence [ADM2, ADM2, ADM1], scale-sequence [5, 5, 7], prominence-sequence [4, 4, 7];
    
    \item specific representation (answer): [San Francisco, Marin County, U.S. state of California]
\end{itemize}

\begin{minipage}{0.9\linewidth}
    \begin{lstlisting}[captionpos=b, caption={SPARQL query of Example 4 (DBPedia)}, label={lst:sparql2},
    basicstyle=\scriptsize \ttfamily,frame=single]
    PREFIX dbo: <http://dbpedia.org/ontology/>

    SELECT distinct ?q1 ?a1 ?a2 ?a3 WHERE {
        VALUES ?q1 {<http://dbpedia.org/resource/Golden_Gate_Bridge>} 

        ?a1 a dbo:PopulatedPlace .
        {?a1 ?r1 ?q1} UNION {?q1 ?r1 ?a1} .

        ?a2 a dbo:PopulatedPlace .
        {?a2 ?r2 ?q1} UNION {?q1 ?r2 ?a2} .

        ?a3 a dbo:PopulatedPlace .
        {?a3 ?r3 ?q1} UNION {?q1 ?r3 ?a3} .
    }
    \end{lstlisting}
\end{minipage}

\begin{minipage}{0.9\linewidth}
    \begin{lstlisting}[captionpos=b, caption={SPARQL query of Example 4 (Geonames)}, label={lst:sparql2_g},
    basicstyle=\scriptsize \ttfamily,frame=single]
    PREFIX gn: <http://www.geonames.org/ontology#>

    SELECT distinct ?q1 ?a1 ?a2 WHERE {
        VALUES ?q1 {<http://sws.geonames.org/5352844/>} 

        ?a1 gn:featureCode gn:A.ADM2 .
        {?a1 ?r1 ?q1} UNION {?q1 ?r1 ?a1} .

        ?a2 gn:featureCode gn:A.ADM2 .
        {?a2 ?r2 ?q1} UNION {?q1 ?r2 ?a2} .


        ?a3 gn:featureCode gn:A.ADM1 .
        {?a3 ?r3 ?q1} UNION {?q1 ?r3 ?a3} .

    }
    \end{lstlisting}
\end{minipage}

\begin{table}\centering
\caption{\label{tab:eg2}SPARQL results to Example 4}
\resizebox{0.7\textwidth}{!}{
    \begin{tabular}{lllll}
    \toprule
        \textbf{Knowledge Base} & \textbf{Q1}      & \textbf{A1}       & \textbf{A2}               & \textbf{A3}           \\
    \midrule
        DBPedia        & Golden Gate Bridge & \cellcolor{green!25}\textbf{San Francisco} & \cellcolor{green!25}\textbf{Marin County} & \cellcolor{green!25}\textbf{California} \\
        
        \midrule
        
        Geonames       & Golden Gate Bridge & \cellcolor{green!25}\textbf{San Francisco} & \cellcolor{red!25}\textbf{--} & \cellcolor{green!25}\textbf{California}\\
    \bottomrule
    \end{tabular}
    }
\end{table}

\section{Feature Codes}
\label{appendix:a}
Table~\ref{tab:feature_codes} shows the types which are mentioned in this paper. The complete list can be found in the Geonames website.

\begin{table}[h!]
	\caption{\label{tab:feature_codes}Feature codes used in the paper (extracted from Geonames documentation)}
	\begin{tabular}{lll}
	\toprule
		\textbf{Code} & \textbf{Description} & \textbf{Example}\\
	\midrule
        ADM1 & first-order administrative division (states, and provinces) & Oklahoma \\
        ADM2 & second-order administrative division (counties) & Brevard County \\
        ADM3 & third-order administrative division (cities) & City of Alhambra \\
        ADM4 & fourth-order administrative division (towns) & Newburgh \\
        AREA & a part of land without homogeneous character/boundaries & Theresienwiese \\
        BDG & a bridge & Putney Bridge \\
        FRM & a part of land dedicated to agricultural purposes & Branksome \\
        HTL & hotels & The Carriage House \\
        MT & mountains & Eagles Nest \\
        PCLI & independent political entity & Paraguay \\
        PPL & diverse type of populated places (e.g., cities, and villages) & El Granada \\
        PPLA2 & seat of a second-order administrative division & Lake City\\ 
        PRK & parks and recreational places & Franklin Square Park \\
        RGN & an area with particular cultural character & Central Africa \\
        SCH & schools and universities & Stuyvesant High School \\
        STM & streams & Withlacoochee River \\
    \bottomrule
	\end{tabular}
\end{table}